# Modeling the Heterogeneous Duration of User Interest in Time-dependent Recommendation: A Hidden Semi-Markov Approach

Haidong Zhang, Wancheng Ni, Xin Li, *Member, IEEE,* and Yiping Yang

*Abstract*—**Recommender systems are widely used for suggesting books, education materials, and products to users by exploring their behaviors. In reality, users' preferences often change over time, leading to studies on time-dependent recommender systems. However, most existing approaches that deal with time information remain primitive. In this paper, we extend existing methods and propose a hidden semi-Markov model to track the change of users' interests. Particularly, this model allows for capturing the different durations users may stay in a (latent) interest state, which can better model the heterogeneity of user interests and focuses. We derive an EM algorithm to estimate the parameter of the framework and predict users' actions. Experiments on three real-world datasets show that our model significantly outperforms the state-of-the-art time-dependent and static benchmark methods. Further analyses of the experiment results indicate that the performance improvement is related to the heterogeneity of state durations and the drift of user interests in the dataset.**

Abstract for Correspondence—Haidong Zhang is pursuing his doctoral degree at the Institute of Automation, Chinese Academy of Sciences. His current research interests include recommender systems and information filtering. He has published several papers in conference proceedings and refereed journals.

| List of Footnotes |
|---|
| Manuscript received .... This work was supported in part by the National Natural Science Foundation of China under Grant 61174190 and 71572169, in part by the GuangDong Natural Science Foundation under Grant 2015A030313876, and in part by CityU SRG 7004287.<br>H. Zhang, W. Ni and Y. Yang are with the Institute of Automation, Chinese Academy of Sciences, Beijing 100190, China (e-mail: haidong_zhang14@yahoo.com; wancheng.ni@ia.ac.cn; yiping.yang@ia.ac.cn).<br>X. Li is with the Department of Information Systems, City University of Hong Kong, Hong Kong (e-mail: Xin.Li.PhD@gmail.com). |

| List of Figures |
|---|
|  |

## I. Introduction

Recommender systems provide personalized suggestions to alleviate the information overload problem of platform users, such as in e-commerce platforms [6], online education platforms [3], and social networking services [9]. They have attracted a lot of attention from industry and academia. Recently, there has been a significant increase in methods capturing contextual information (e.g., location, time, device, etc.) to improve the performance of recommender systems [12], [15], [17].

Time is a strong indicator of context. In recommender systems, people's preferences and the environment often evolve over time [19], [22]. For example, in the tourism domain, users' preferred travelling destinations vary across seasons [19]. In music and movie consumption, users' choices may be affected by their mood, which is different on weekdays and weekends [20], [21]. Time-dependent recommender systems that can track the evolution of users' preferences over time is an important area of recommender systems [19].

Some straightforward options exist to apply and vary mature methods to deal with time. For example, one can identify sequential consumption patterns of items [2] or contents [4] to infer future consumption. One can apply a time-decay factor to weight the importance of historical consumption to convert a time-independent model to a time-dependent model [8], [16], [29]. The third method is to design features related to time (e.g., users' periodic consumption, seasonal factors) as predictors [20], [23]. While these methods provide improvements to time-independent models, they are limited by their mechanisms. The sequence mining method and time-decay method ignore or over-simplify the impact of time on human behavior. The time-related feature method essentially converts the difficulty of modeling to the difficulty of feature design, which highly depends on researchers' domain knowledge and is labor intensive.

A recent approach in time-dependent recommender systems is to model user's interest drift over time as the transition of latent states in a hidden Markov model (HMM) framework. Sahoo*, et al.* [28] presented a prominent work on this approach by combining HMM with the aspect model. The aspect model models user interests as the distributions of consumptions. The HMM models transitions of each user's interest and captures heterogeneity of user interests over time. It achieves a moderate performance improvement over existing methods. But it provides a generic and insightful framework for time-dependent recommendation.

While significantly expanding the modeling of time-related factors, the HMM leaves some room for theoretical improvements and possible performance improvements. Particularly, the HMM setup assumes users have a probability to leave a state at the end of its duration. But the probability of staying in one state for multiple periods decreases exponentially, and the duration for users' stay in different interest states follows a geometric distribution. This assumption may not fit the reality of all users in all contexts. Thus, in this paper, we improve on Sahoo, *et al.* [28] by proposing a more flexible framework to capture the heterogeneity in the duration of users' interests for time-dependent recommendation.

Specifically, we build a hidden semi-Markov model (HSMM) framework to tackle users' drifting interests. To the best of our knowledge, this is the first work that introduces a hidden semi-Markov model to address the recommendation problem. The model captures users' heterogeneity in interest duration by allowing them to stay in different (latent) states for different time periods, which is modeled in a semi-parametric manner. We derive an expectation maximization (EM) algorithm to estimate the model parameters and predict users' interests. Since users' multiple period stay in one state is explicitly modeled, the prediction also considers multiple periods' data (in a semi-Markov fashion). We employ three real-world datasets to evaluate the proposed model and inspect the existence of heterogeneous interest duration. Experiments show that our model is more effective than both the classic static algorithms and the state-of-the-art time-dependent methods, including the HMM [28].

The main contributions of this paper are three-fold. First, we explicitly explore a new dimension of time-related factors in recommendation: the duration of user interest. Through the experiments, we identify the existence of heterogeneous interest durations, which is seldom reported in literature. Second, we develop a generic framework combining HSMM with the aspect model for time-dependent recommendation. The model can model heterogeneous user interest durations together with the changing of user interests. The deduction of the model and the EM algorithm is not trivial and has significant practical value for recommender system implementations. Third, we conduct a comprehensive evaluation of our proposed framework as compared with classic static methods and the state-of-the-art time-dependent methods. The high prediction effectiveness of our model as shown from three real-world datasets indicates a good potential for its application in time-dependent recommender system applications.

## II. Related work

Since several papers (e.g., [12], [30], [31]) have conducted comprehensive reviews on recommender systems, we focus our review on time-dependent recommender systems [19], [28] that leverage the time information in recommendation. According to our review, there generally exist four approaches for time-dependent recommendation.

The first method is to identify sequential patterns of consumption from items' access history. Future item consumption is predicted as the items associated with historical items. For example, Géry and Haddad [1] and Awad and Khalil [2] discovered sequential patterns from web logs with the Apriori algorithm for webpage recommendation. Huang, *et al.* [3] modeled the transition probabilities among education materials as Markov chains and made recommendations based on the learning paths of users. In music recommendation, Hariri, *et al.* [4] identified topics of songs and then discovered the sequential patterns of topics with the Generalized Sequential Pattern mining algorithm for inferring the next topics most appealing to users. Zhang and Chow [5] utilized Markov chains to represent the transition of points-of-interest (POIs) in tourism site recommendation and predict the next POI that may be visited by a user. Chen, *et al.* [7] applied sequential pattern mining after item-based CF to filter the recommended items and improve prediction performance.

The second method is to employ a time window or apply a decay factor on the time variable in classic recommendation models. This approach is widely adopted due to its simplicity. In a time-window method, only activities within the time are considered. For example, Xiang, *et al.* [18] split data records according to time and consider records in a time window as a session to model users' preferences. In a decay factor method, recent activities contribute more weight to users' current interests. A common practice is to reduce the weight of ratings on items over time in item-based collaborative filtering (CF) [8], [10], [11], [14]. Basile, *et al.* [13] built a time-adaptive user profile that combined decay factors and the content-based similarity of items. In model-based collaborative filtering, Dunlavy, *et al.* [16] built a time-decay matrix under a singular vector decomposition (SVD) framework to increase weights of recently browsed items.

The third method is to develop time-related variables for machine learning models. The time dimension has different meanings in different contexts [19], [25] and may lead to different feature designs. For example, Baltrunas and Amatriain [20] mainly discussed building time variables based on their semantics, such as morning vs. evening, weekday vs. weekend, cold season vs. hot season, etc. Bogina, *et al.* [26] conducted an ensemble of two clustering methods, Bagging classifier and Naive Bayes Tree, on the number of consumptions in different time scales, such as month of the year or day of the week. Xiang and Yang [21] proposed a TimeSVD, which considers the change of user and item bias as a day/month/year-based variable in the matrix factorization framework. Koren [23] took a similar approach by incorporating time-variant item bias based on season and bin (10 weeks) and time-variant user bias based on season and day into the matrix factorization method to capture the changing characteristics for user and item. Ramirez-Garcia and Garcia-Valdez [24] built multiple rating matrices by segmenting time context sequentially, and combined user-based CF and the fuzzy inference system considering user's participation and item popularity for recommendation.

TABLE I
TIME-EVOLVING RECOMMENDATION APPROACHES

| **Studies** | **Model & Algorithm** | **Usage of time** | **Application** |
|---|---|---|---|
| Sequence pattern mining (SPM) | | | |
| Géry and Haddad [1] | Apriori algorithm | Time as an annotation of access sequences | Webpage |
| Awad and Khalil [2] | Apriori algorithm + Markov chain model | | Webpage |
| Huang, *et al.* [3] | Markov chain model + Entropy-based paths ranking | | Education material |
| Hariri, *et al.* [4] | Generalized sequence mining (on clusters of items) | | Music |
| Zhang and Chow [5] | Markov chain model Kernel density estimation | | Tourism |
| Chen, *et al.* [7] | Item-based CF Generalized sequence mining | | Education material |
| Time Window/Time decay | | | |
| Ding and Li [8] | Item-based CF | Weighting terms according to access recency | Movie |
| Liu, *et al.* [10] | Clustering, Content-based, Item-based CF | | Movie |
| Yu and Li [11] | Item-based CF | | Movie |
| Basile, *et al.* [13] | Content-based | | Movie |
| Su, *et al.* [14] | Item-based CF | | Movie |
| Dunlavy, *et al.* [16] | SVD, Katz | Time-decay matrix | Document |
| Xiang, *et al.* [18] | Session-based Temporal Graph | Splitting data records according to time | Document; Webpage |
| Time-related features | | | |
| Baltrunas and Amatriain [20] | Matrix factorization | Variables defined on day, week, month, season, year, etc. | Music |
| Xiang and Yang [21] | timeSVD | | Movie |
| Koren [23] | timeSVD++ | | Movie |
| Ramirez-Garcia and Garcia-Valdez [24] | User-based CF, Item popularity | | Movie |
| Darapisut, *et al.* [25] | Rules-based tendencies | | Music |
| Bogina, *et al.* [26] | Bagging classifier, Naive Bayes Tree | | E-commerce |
| Transitive latent variables | | | |
| Xiong, *et al.* [27] | Probabilistic Tensor Factorization | The transition of latent states along with time | Products; Movie |
| Sahoo, *et al.* [28] | HMM+Aspect model | | Blog; Movie; Music |

The fourth method is to model users' shifting interests with transitive latent variables. It assumes users' interests in each time period are determined by their latent states. Xiong, *et al.* [27] proposed a Bayesian Probabilistic Tensor Factorization that modeled time-stamped user-item ratings as the inner product of three dimensional vectors: user factors, item factors, and time factors of depending on the preceding time step. Sahoo, *et al.* [28] incorporated the aspect model into HMM for tracking users' interest changes with the transition of latent states.

Table 1 summarizes the prior time-dependent recommender system studies based on the four approaches. We observe that sequence, time window/decay factor, and customized time-based variables are the three major approaches taken by previous studies. Such methods extend static methods. They also have their inherent limitations which restrict their applications:

1) The sequential pattern mining methods consider only access sequences of items. They essentially leverage sequence-related rather than time-related information. The same purchase sequence may appear in totally different time scopes and have significantly different user behavior implications.
2) The time window/decay factor methods apply arbitrary rules to weight the relationship between users' historical activities and their future activities (i.e., weight as 0/1 or a decaying number). Such rules create a very strong assumption of people's behavior and cannot capture the different speeds of interest drift over time.
3) The methods based on time-related features essentially convert the difficulty of modeling to the difficulty of feature design. This practice highly depends on the application context and requires significant domain knowledge and effort to develop, which limits their generalizability.

In this study, we favor the fourth approach of transitive latent state methods, since they bring us a new perspective in modeling time. By combining HMM with the aspect model, the time effect is converted to the transition of states in Sahoo*, et al.* [28], which allows both a stable user interest (or small interest drift) within each state (in each time period) and a changing user interest across states (across time periods). The model captures the heterogeneity of users by allowing them to be in different states at a time period. Essentially, the model allows leveraging a different period of historical data (or weighting historical data differently) for each user to predict the user's future interest. Nevertheless, the HMM setup implies a geometric distribution of state duration [32], [33], where the probability of staying in one state for multiple periods decreases exponentially with increased duration. This implication is different from users' behavior when the time period is reduced to a shorter level, since some users may have a relatively stable interest last for multiple time periods. The assumption may limit the prediction performance.

In this research, we take the fourth approach in time-dependent recommendation and aim to tackle the problem of modeling the heterogeneity of user interest durations.

## III. A Hidden Semi-Markov Framework for Time-dependent Recommendation

### A. *System Framework*

Our model extends [28] and combines the hidden semi-Markov model (HSMM) with the aspect model to make predictions. According to our literature review, the HSMM approach has not been explored in recommendation literature. We explicitly allow latent states lasting for multiple time periods to capture the heterogeneity in users' interest durations. We derive an EM algorithm with maximum a posteriori (MAP) estimation to implement the framework [34], [35], [36].

Fig. 1 presents the general process of our proposed method. In this framework, recommendation is conducted by predicting the number of consumptions of each item by each user in a certain period (e.g., a month, a week, etc.). Thus, at the data preparing stage, we aggregate raw transactions to the monthly level.

The aggregation provides user-item interactions/consumptions for each month, which are used as training and testing data.

At the model training stage, the matrix of user-item consumptions is used to estimate the model parameters. The model details are reported in the next subsection. In general, the model captures the probability of total item consumptions in a period and the probability distributions of consuming different items. The two probabilities are affected by the interest state of users, which is modeled as an HSMM in this study. The HSMM specifies the state transition relationship and state duration probabilities for tracing each user's shifting interest. We develop an EM algorithm to estimate the parameters of the model.

At the prediction stage, we first infer each user's next latent state based on the transition and duration of states using HSMM. After specifying the state, the top-N recommendations can be identified based on the corresponding probability distributions of items and possible total consumptions. In this study, prediction is conducted immediately after training the data.

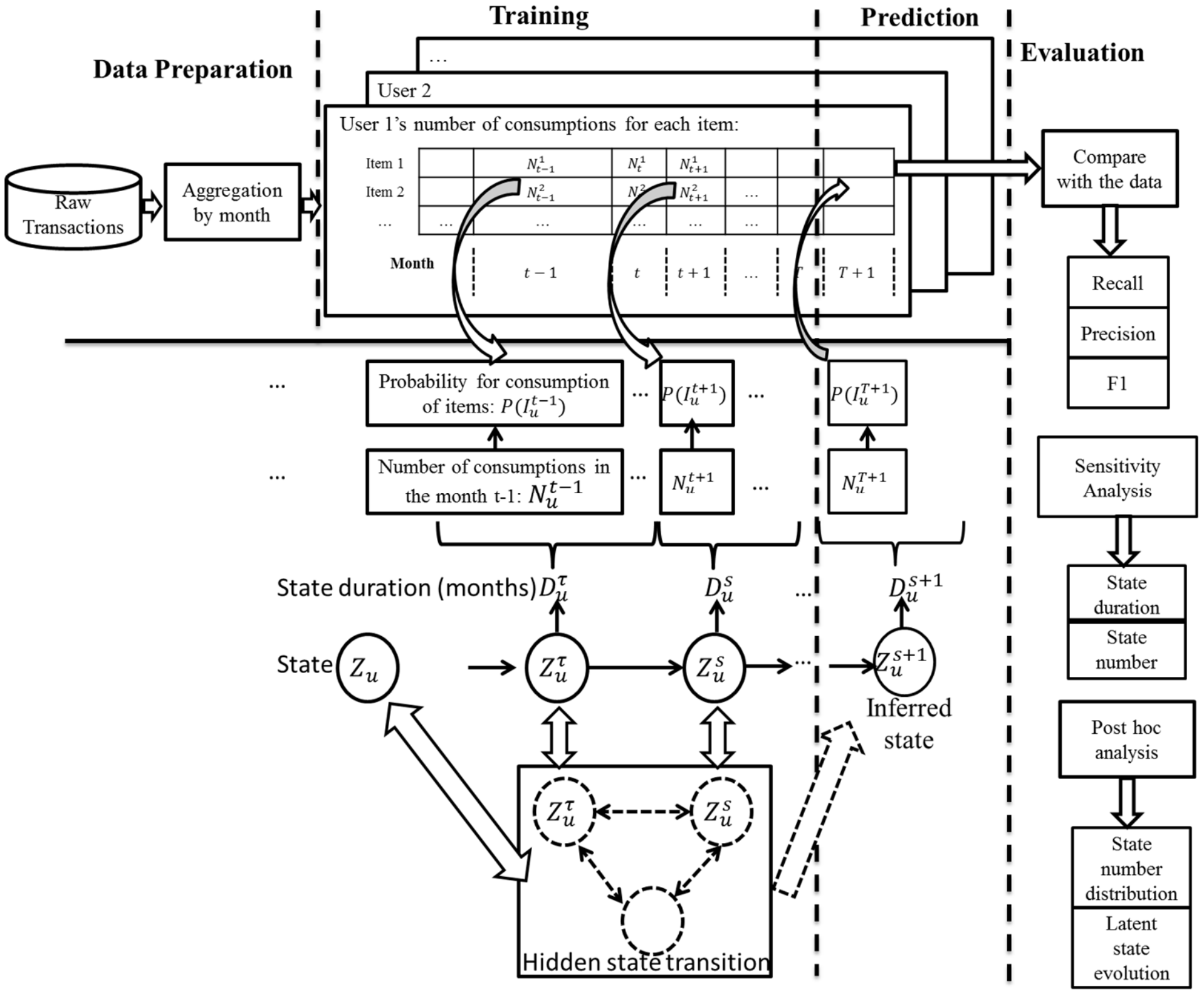

Fig. 1. System framework of the hidden semi-Markov model for collaborative filtering

At the evaluation stage, the predicted numbers of consumptions are compared with the real consumptions in the prediction period. Prediction performance is evaluated by precision, recall, and F-measure. To understand the characteristics of our proposed model, in this study we analyze the performance of our model according to the change of state duration and state number. We also conduct analysis on the state duration distributions and latent state evolutions in the experimental datasets.

## B. *Model Setup*

Our proposed framework is based on HSMM, which is an extension of the HMM. In both HMM and HSMM, observations of the item consumptions are considered emitted from the hidden states (one observation in each time period). Similar to HMM, HSMM assumes the state transition follows a Markov characteristic, i.e., the future states are conditionally independent of historical states given the current state [37], [33]. Different from HMM, HSMM allows each state to have a variable time duration that can last multiple time periods [18]. So at state level, the HSMM can emit multiple observations. At individual time period level, the underlying stochastic process of HSMM model is semi-Markov. HSMM has been widely used to deal with sequential data and time- related data, such as speech recognition [38], handwriting recognition [39], and gene sequence analysis [40].

To model the heterogeneity of user interest durations, our model considers the state duration following a non-parametric distribution. We also specify the self-transition probability for each state to be zero, i.e., a state would be left at the end of its duration. Both setups are classic in HSMM [33], [41]. Fig. 2 shows the graphical representation of our proposed model. Assume a dataset with $|U|$ users consuming items in $T$ time periods, $t \in [1, T]$. Each user follows the same structure of shifting interests. We assume the possible interest of users with a set of latent states $S = \{1,2, \dots, K\}$, and their initial probabilities are $\pi$. A discrete latent variable $Z_u^\tau \in S$ is employed to denote the $\tau$-*th* segmental state for user *u* in the dataset, which is similar to the aspect model [42]. The changes of users' interests are captured by the transition probability between states, represented with $A$, as in Sahoo, *et al.* [28]. We use $D_u^\tau$ to represent the duration of state $Z_u^\tau$, which lasts from time period $t - D_u^\tau + 1$ to $t$. The probability of state duration is represented as *D*. In each of the $D_u^\tau$ time periods, say, at time *t*, the user consumes a set of items $I_u^t$ for a total of $N_u^t$ times, which are the emissions of the model. They separately follow multinomial distributions with parameters $\theta$ and negative binomial distributions with parameters $(r, p)$. Next, we elaborate and justify our model specification.

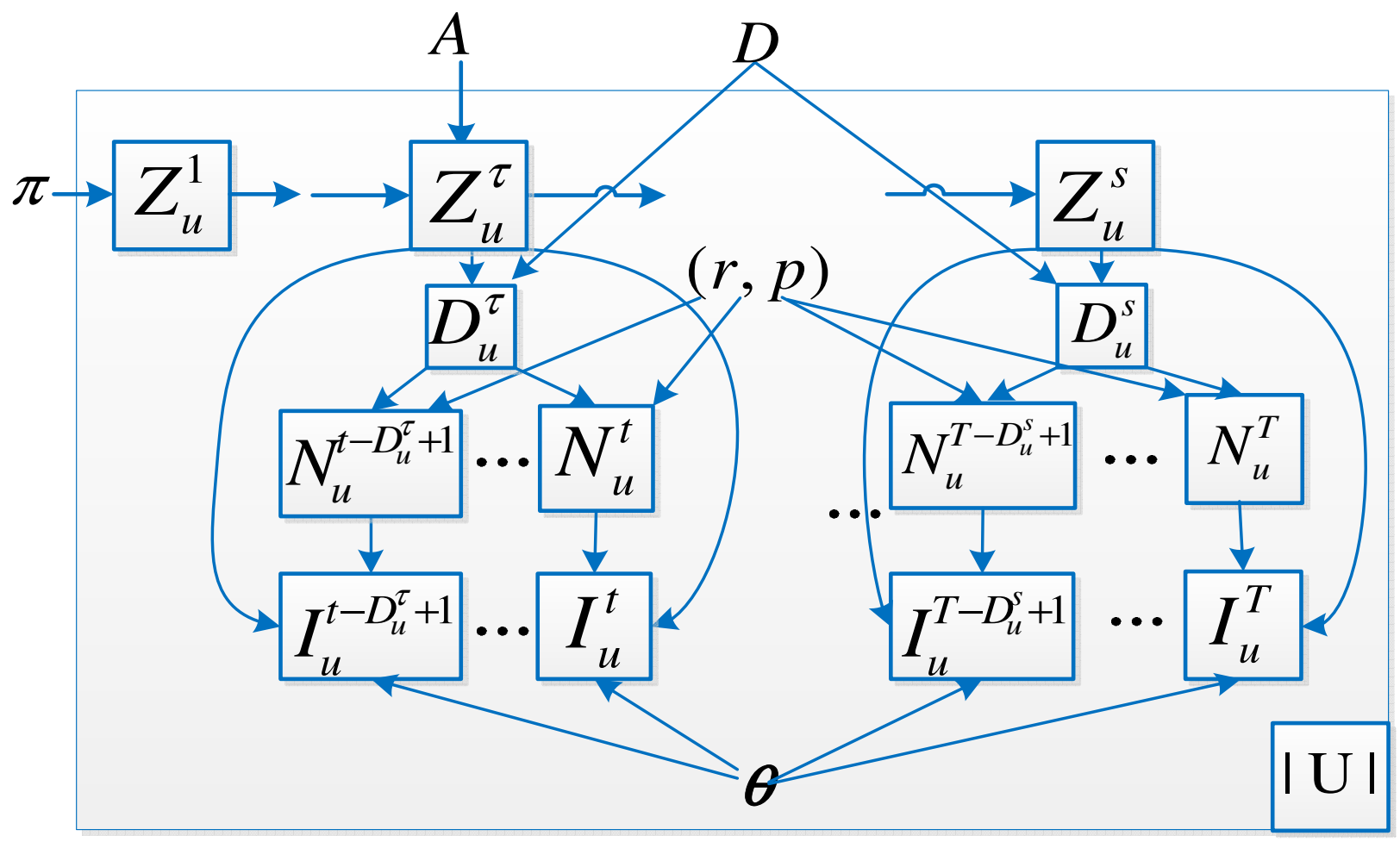

Fig. 2. A hidden semi-Markov model for collaborative filtering.

1. **State transition**. In our model, the $\tau$-th segmental state is represented as $Z_u^\tau = i$, which starts at $t - D_u^\tau + 1$ and ends at *t* with duration $d$, as $(Z_u^t = i, D_u^t = d)$. The segmental state transition probability from the ($\tau$-*1*)-th segment to the $\tau$-th segment is:

$$P(Z_u^\tau = j \mid Z_u^{\tau\text{-}1} = i) = P(Z_u^t = j, D_u^t = d \mid Z_u^{t\text{-}d} = i, D_u^{t\text{-}d} = d') \qquad (1)$$

Since there are $K$ states and each state can last for $M$ periods, the size of the state transition matrix is $KM * KM$. In order to simplify the model, we assume the segmental state transition probability is independent from the state duration probability given the current state [43], i.e.:

$$\begin{aligned} & P(Z_{\text{u}}^t = j, D_{\text{u}}^t = d \mid Z_{\text{u}}^{t-d} = i, D_{\text{u}}^{t-d} = d') \\ & = P(Z_{\text{u}}^t = j, D_{\text{u}}^t = d \mid Z_{\text{u}}^{t-d} = i) \\ & = P(Z_{\text{u}}^t = j \mid Z_{\text{u}}^{t-d} = i) P(D_{\text{u}}^t = d \mid Z_{\text{u}}^t = j) \\ & = \mathrm{A}_{i,j} D_{j,d} \end{aligned} \qquad (2)$$

This design allows us to decompose segmental state transition to the multiplication of $A_{i,j} = P(Z_u^t = j|Z_u^{t-d} = i)$ at the state level, and $D_{j,d} = P(D_u^t = d|Z_u^t = j)$ at the state duration level. Note that, $\sum_j A_{i,j} = 1$ $(i \neq j)$ and $\sum_d D_{j,d} = 1$.

One major difference between HSMM and HMM is modeling the state duration. Since we do not have prior knowledge on the distribution of state duration, we take a non-parametric approach and learn this distribution *D* from the data. This is a classic setup in HSMM [41], [32]. When taking this non-parametric approach, it is necessary to specify an upper bound for state duration to control computational complexity, which is noted with *M* in this paper. So, the interests of users can last at most *M* time periods. With the zero probability of self-transition defined in the classic HSMM, users will leave each interest state when it ends.

2. **Emission (i.e., item consumption).** The emission process defines the probabilities of the number of items consumed and the distributions of items consumed. We assume the dataset has $|I|$ items, and the number of consumptions of the $i$-th element is $x_i$. The set of items consumed by user *u* at time *t* can be represented as $I_u^t = [x_1, x_2, \dots, x_{|I|}]_u^t$. So, the total number of views is $N_u^t = \sum_i x_i$. We assume $N_u^t$ follows a set of negative binomial distributions (NBD). The NBD distribution is often chosen to model the count data due to its fitting to real-world data distributions[1]. For example, it can provide a highly skewed distribution [44]. It also allows the variance to be greater than its mean and can overcome the overdispersion problem that occurs when using Poisson distribution to model count data [45]. Previously, many studies in e-commerce [28], [46] employed NBD to model the number of consumptions. In our model, we specify the NBD distribution of $N_u^t$ as:

$$P(N_u^t; r_{k,d}, p_{k,d}) = \frac{\Gamma(N_u^t + r_{k,d})}{N_u^t!\,\Gamma(r_{k,d})} p_{k,d}{}^{N_u^t} (1 - p_{k,d})^{r_{k,d}} \quad (3)$$

where $r_{k,d}$ and $p_{k,d}$ are the parameters given a state *k* lasting for *d* time periods.

[1] For this study, we compared different distributions on our datasets and found that NBD better fits our data.

In addition to the total number of consumptions, we need to model distributions of consumptions on different items. As in the research on HMM and aspect models [28], we assume users' preferences for items are conditionally independent of their latent states. Hence, the item consumptions can be considered as $N_u^t$ independent Bernoulli trials, which follow multinomial distributions. We specify this distribution as:

$$P(\mathrm{I}_{\mathrm{u}}^{t};\theta_k)=\frac{N_{\mathrm{u}}^{t}!}{x_1!\ldots x_{|I|}!}\theta_{k,1}^{x_1}\ldots\theta_{k,|I|}^{x_{|I|}} \tag{4}$$

where $\sum_{i=1}^{|I|} x_i = N_u^t$, $\sum_{i=1}^{|I|} \theta_{k,i} = 1$. $\theta$ is a parameter matrix. $\theta_{k,i}$ is the probability of the *i-th* item to be selected given state $k$.

Overall, the probability for the $\tau$-th state to emit a set of items from time $t - D_u^\tau + 1$ to $t$ is:

$$P(I_u^{t-d+1:t} \mid Z_u^{t-d+1:t}=\mathrm{i})=\prod_{\tau=t-d+1}^{t} P(N_{\mathrm{u}}^{\tau},\mathrm{I}_{\mathrm{u}}^{\tau};r_{i,d},p_{i,d},\theta_i)=\prod_{\tau=t-d+1}^{t} P(N_{\mathrm{u}}^{\tau};r_{i,d},p_{i,d})P(\mathrm{I}_{\mathrm{u}}^{\tau};\theta_i) \tag{5}$$

3. **The initial state distribution.** Similar to the specification of state transition probability, we decompose state and state duration and specify the initial distribution as:

$$\mathrm{P}\left(\mathrm{Z}_{\mathrm{u}}^{1:\mathrm{d}}=k\right)=P\left(Z_u^1=k,D_u^t=d\right)=P\left(Z_u^1=k)P(D_u^t=d \mid Z_u^1=k\right)=\pi_k D_{k,d} \tag{6}$$

where $\pi_k$ is the initial distribution of state $k$, and $\sum_k \pi_k = 1$.

In general, by combining the state transition, emission, and initial state distribution, our model can be specified by the following parameters:

- $\pi$: The initial probability of the states.
- $A$: The transition probability between states.
- $D$: The state duration probability.
- *r, p*: Parameters of NBD followed by $N_u^t$.
- $\theta$: Parameter of multinomial distribution for $I_u^t$.

After specifying these parameters, we are able to infer the (latent) state of each user at each time period and the possibility of the user consuming an item in the time period. In subsection III-C, we present an EM algorithm to estimate the parameters. In subsection III-D, we present the inference algorithm that predicts the consumption probability.

## C. *EM Algorithm*

According to our specification above, the joint distribution of observing items $\{I_u^{1:T}\}_{|U|}$ selected by each user $u$ from the first time period to the *T-th* time period in the training part of the dataset is:

$$\begin{aligned} p(Z_u^t, I_u^{1:T} \mid \Theta) &= \prod_{k=1}^{K}\prod_{d}^{M} p(Z_u^{1:d} = k \mid \pi)\prod_{t=2}^{T}\prod_{j\neq k}^{K}\prod_{k}^{K}\prod_{\mathrm{d}}^{M} p(Z_u^{t-d+1:t} = k \mid Z_u^{t-d} = j, A)\prod_{t=1}^{T}\prod_{k}^{K} p(\{I_{uj}^t\} \mid Z_u^t = k, \theta, r, p) \\ &= \prod_{k=1}^{K}\prod_{d}^{M} p(Z_u^1 = k \mid \pi)\mathrm{D}_{k,d}\prod_{t=2}^{T}\prod_{j\neq k}^{K}\prod_{k}^{K}\prod_{\mathrm{d}}^{M} p(Z_u^t = k \mid Z_u^{t-d} = j, A)\mathrm{D}_{k,d}\prod_{t=1}^{T}\prod_{k}^{K} p(\{I_{uj}^t\} \mid Z_u^t = k, \theta, r, p) \qquad (7) \\ &= \prod_{k=1}^{K} p(Z_u^1 = k \mid \pi)\prod_{t=2}^{T}\prod_{j\neq k}^{K}\prod_{k}^{K}\prod_{\mathrm{d}}^{M} p(Z_u^t = k \mid Z_u^{t-d} = j, A)\prod_{t=1}^{T}\prod_{k}^{K}\prod_{\mathrm{d}}^{M}\mathrm{D}_{k,d}\prod_{t=1}^{T}\prod_{k}^{K} p(\{I_{uj}^t\} \mid Z_u^t = k, \theta, r, p) \end{aligned}$$

where $I_u^{1:T}$ represents the observations spanning over $T$ time periods for user $u$.

We adopt an expectation maximization framework based on the forward-backward algorithm to find the maximum a posteriori (MAP) estimation of the parameters of our model given the observations from the training data. According to Bayesian theory, the MAP estimation incorporates prior distributions over the model parameters [47]. We express the prior knowledge by assuming the parameters of prior distributions to be fixed, which can alleviate the overfitting problem and reduce the complexity of the training process [35], [48], [28]. Following the EM framework, we alternate between computing the forward and backward variables to derive the posterior distributions given the observations in the E step and updating the model parameters under the MAP criteria in the *M* step until the joint likelihood converges.

The log likelihood given old parameter $\Theta^{\text{old}}$ is:

$$
\begin{aligned}
Q(\Theta,\Theta^{old}) &= \sum_{u}\sum_{k=1}^{K} p(\mathrm{Z}_u^1 = k \mid \mathrm{I}_\mathrm{u}^{1:\mathrm{T}};\Theta^{old}) \log \pi_\mathrm{k} \\
&+ \sum_{u}\sum_{t=2}^{T}\sum_{j\neq k}^{K}\sum_{k}^{K} p(Z_u^{t-1} = j,\ Z_u^t = k \mid \mathrm{I}_\mathrm{u}^{1:\mathrm{T}};\Theta^{old}) \log \mathrm{A}_{\mathrm{jk}} \\
&+ \sum_{u}\sum_{t=1}^{T}\sum_{k}^{K}\sum_{\mathrm{d}}^{M} p(Z_u^t = k, D_u^t = d \mid \mathrm{I}_\mathrm{u}^{1:\mathrm{T}};\Theta^{old}) \log \mathrm{D}_{k,d} \\
&+ \sum_{u}\sum_{t=1}^{\mathrm{T}}\sum_{k}^{K}\sum_{\mathrm{d}}^{M} [p(Z_u^t = k, D_u^t = d \mid \mathrm{I}_\mathrm{u}^{1:\mathrm{T}};\Theta^{old}) \log p(N_\mathrm{u}^\mathrm{t} \mid \mathrm{r}_{k,d}, p_{k,d})] \\
&+ \sum_{u}\sum_{t=1}^{T}\sum_{k}^{K} p(\mathrm{Z}_\mathrm{u}^t = k \mid \mathrm{I}_\mathrm{u}^{1:\mathrm{T}};\Theta^{old}) \log p(\{I_{uj}^t\} \mid \theta_k, N_\mathrm{u}^\mathrm{t})
\end{aligned}
\tag{8}
$$

To estimate the parameters ($\pi$, $A$, $D$, $r$, $p$, and θ) listed in formula (8), we follow the forward-backward algorithm that computes the posterior probabilities. The forward variables and the backward variables are:

$$\alpha(Z_u^t = \mathrm{i}, D_u^t = \mathrm{d}) = P(Z_u^{t-d+1:t} = \mathrm{i} \mid I_u^{1:\mathrm{t}}) \tag{9}$$

$$\beta(Z_u^t = \mathrm{i}, D_u^t = \mathrm{d}) = \mathrm{P}(I_u^{t+1:\mathrm{T}} \mid Z_u^{t-d+1:t} = \mathrm{i}) / P(I_u^{t+1:\mathrm{T}} \mid I_u^{1:\mathrm{t}}) \tag{10}$$

Note that the forward and backward variables can be recursively calculated as follows (Appendix A):

$$\alpha(Z_u^t = \mathrm{i}, D_u^t = \mathrm{d}) = [\sum_{j\in \mathrm{S}\setminus\{\mathrm{i}\}}\sum_{d'} \alpha(Z_u^{t-d} = j, D_u^{t-d} = \mathrm{d'}) A_{ji}] P(I_u^{t-d+1:\mathrm{t}} \mid Z_u^{t-d+1:t} = \mathrm{i}) D_{i,d} / \mathrm{P}(I_u^{\mathrm{t-d+1:t}} | I_u^{1:\mathrm{t-d}}) \tag{11}$$

$$\beta(Z_u^t = \mathrm{i}, D_u^t = \mathrm{d}) = \sum_{j\in S\setminus\{i\}}\sum_{d'} [\beta(Z_u^{t+d'} = j, D_u^{t+d'} = \mathrm{d'}) \mathrm{P}(I_u^{t+1:t+d'} \mid Z_u^{t+1:t+d'} = \mathrm{j}) A_{ij} D_{j,d'} / P(I_u^{t+1:\mathrm{t+d'}} \mid I_u^{1:\mathrm{t}})] \tag{12}$$

where $P(I_u^{t-d+1:t}|I_u^{1:t-d})$ is included to normalize the variables so that $\sum_i \alpha(Z_u^t = i, D_u^t = d) = 1$.

After determining the forward and backward variables, we can estimate the posterior distributions in formula (8). Firstly, the posterior of $\mathrm{P}(Z_u^t = i, D_u^t = d | I_u^{1:T})$ is the product of α and β.

$$
\begin{aligned}
& P(Z_u^t = \mathrm{i}, D_u^t = \mathrm{d} \mid I_u^{1:\mathrm{T}}) = P(Z_u^{t-d+1:t} = \mathrm{i} \mid I_u^{1:\mathrm{T}}) \\
&= P(I_u^{1:T} \mid Z_u^{t-d+1:t} = \mathrm{i}) P(Z_u^{t-d+1:t} = \mathrm{i}) / P(I_u^{1:\mathrm{T}}) \\
&= \frac{P(I_u^{1:\mathrm{t}} \mid Z_u^{t-d+1:t} = \mathrm{i}) P(Z_u^{t-d+1:t} = \mathrm{i})}{P(I_u^{1:\mathrm{t}})} \frac{\mathrm{P}(I_u^{t+1:\mathrm{T}} \mid Z_u^{t-d+1:t} = \mathrm{i})}{P(I_u^{1:\mathrm{T}}) / P(I_u^{1:\mathrm{t}})} \\
&= P(Z_u^{t-d+1:t} = \mathrm{i} \mid I_u^{1:\mathrm{t}}) \mathrm{P}(I_u^{t+1:\mathrm{T}} \mid Z_u^{t-d+1:t} = \mathrm{i}) / P(I_u^{t+1:\mathrm{T}} \mid I_u^{1:\mathrm{t}}) \\
&= \alpha(Z_u^t = \mathrm{i}, D_u^t = \mathrm{d}) \beta(Z_u^t = \mathrm{i}, D_u^t = \mathrm{d})
\end{aligned}
\tag{13}
$$

Secondly, the posterior probability of a user staying in state *i* at *t* given all observations, $P(Z_u^t = i|I_u^{1:T})$, is equal to the sum of the probabilities of state *i* with different durations, which is:

$$\begin{aligned}
\mathrm{P}(Z_u^t = \mathrm{i} \mid \mathrm{I}_u^{1:\mathrm{T}}) &= \sum_{d=1}^{M} \mathrm{P}(Z_u^t = \mathrm{i}, D_u^t = \mathrm{d} \mid I_u^{1:\mathrm{T}}) + \sum_{d=2}^{M} \mathrm{P}(Z_u^{t+1} = \mathrm{i}, D_u^{t+1} = \mathrm{d} \mid I_u^{1:\mathrm{T}}) + \ldots \\
&\quad + \sum_{d=M}^{M} \mathrm{P}(Z_u^{t+M-1} = \mathrm{i}, D_u^{t+M-1} = \mathrm{d} \mid I_u^{1:\mathrm{T}}) \\
&= \sum_{s \geq t}^{t+M-1} \sum_{d=\tau-t+1}^{M} \mathrm{P}(Z_u^s = i, D_u^s = \mathrm{d} \mid I_u^{1:\mathrm{T}})
\end{aligned} \tag{14}$$

Note that the state ($Z_u^s = i$, $D_u^s = d$) can also be expressed as $Z_u^{s-d+1:s} = i$. The ranges of sum, $s \geq t$ and $d \geq s - t + 1$, ensure that this user must stay in state $i$, i.e., $Z_u^t = i$.

Thirdly, since the transition of users' interests includes all possible durations of the preceding and succeeding states, the posterior probability that a user changed from state $j$ to state $i$ at time period $t+1$, $P(Z_u^{t+1} = i, Z_u^t = j \mid I_u^{1:T})$, (given $i, j \in S$, $i \neq j$), can be calculated as:

$$P(Z_u^{t+1} = \mathrm{i}, Z_u^t = j \mid \mathrm{I}_u^{1:\mathrm{T}}) = \sum_d \sum_{d'} P(Z_u^{t+1:t+d} = \mathrm{i}, Z_u^{t-d'+1:t} = j \mid I_u^{1:\mathrm{T}}) \tag{15}$$

In this formula, $P(Z_u^{t+1:t+d} = \mathrm{i}, Z_u^{t-d'+1:t} = j \mid I_u^{1:\mathrm{T}})$, the posterior probability that a user's segmental state turning from state $j$ with duration $d'$ to state $i$ with duration $d$ at time period $t+1$ given observations $I_u^{1:T}$ is:

$$\begin{aligned}
&P(Z_u^{t+1:t+d} = \mathrm{i}, Z_u^{t-d'+1:t} = j \mid I_u^{1:\mathrm{T}}; i \neq j) \\
&= \alpha(Z_u^t = j, D_u^t = \mathrm{d'}) P(I_u^{t+1:\mathrm{t+d}} \mid Z_u^{t+1:t+d} = \mathrm{i}) \mathrm{P}(Z_u^{t+1:t+d} = \mathrm{i} \mid Z_u^{t-d'+1:t} = \mathrm{j}) \beta(Z_u^{t+d} = \mathrm{i}, D_u^{t+d} = \mathrm{d}) / \mathrm{P}(I_u^{t+1:\mathrm{t}+d} \mid I_u^{1:\mathrm{t}})
\end{aligned} \tag{16}$$

In addition to the posteriors defined from formulas (9)-(15), formula (8) also depends on multiple initial probabilities, $\pi, A, D, \theta, r,$ and $p$. For the initial state distributions $\pi$, let's use state vector $X = [x_1, x_2, \ldots, x_K]$ to represent whether a user is in each state (1 means in the state, 0 means not in the state). A user's latent state in the first time period ($t=1$) can be considered as a multinomial trials process with one trial. The conjugate prior of multinomial distribution is the Dirichlet distribution. So $\pi$ can be specified as:

$$\pi \sim Dirichlet(x \mid \alpha_1, \ldots, \alpha_K) \tag{17}$$

For the sake of simplicity, here we assume the multiple parameters of the Dirichlet distribution are the same, i.e., $\alpha_i = \alpha / \mathrm{K}$. In our experiments, we set $\alpha$ to 100.

Similarly, the distribution of a user's latent state transition and state duration conditional on each user's latent state are multinomial. We can also specify $A$ and $D$ using Dirichlet distributions as follows:

$$A_{j,:} \sim Dirichlet(x \mid \alpha_1, \ldots, \alpha_K), \text{ where } \alpha_i = \alpha / K \tag{18}$$

$$D_{k,:} \sim Dirichlet(x \mid \alpha_1, \ldots, \alpha_K), \text{ where } \alpha_i = \alpha / M \tag{19}$$

Since the items consumed by a user at time period $t$, $I_u^t = \left[x_1, x_2, \ldots, x_{|I|}\right]_u^t$, conditional on the user's latent state, also follow a multinomial distribution, the distribution of item consumption $\theta_k$ can be specified as a Dirichlet distribution:

$$\theta_{k,:} \sim Dirichlet(x \mid \alpha_1, \alpha_2, \ldots, \alpha_{|I|}), \text{ where } \alpha_i = \alpha / |I| \tag{20}$$

With these specifications, as shown in Appendix B, the MAP of the parameters can be inferred as:

$$\widehat{\pi_k} = \frac{\sum_u P(Z_u^1 = k \mid I_u^{1:T}; \Theta^{old}) + \alpha_k - 1}{\sum_u \sum_k^K P(Z_u^1 = k \mid I_u^{1:T}, \Theta^{old}) + \alpha - K} \tag{21}$$

$$\widehat{A_{jk}} = \frac{\sum_u \sum_{t=2}^T p(Z_u^t = k,\ Z_u^{t-1} = j \mid I_u^{1:T}; \Theta^{old}) + \alpha_k - 1}{\sum_u \sum_{t=2}^T \sum_{k \neq j}^K p(Z_u^t = k,\ Z_u^{t-1} = j \mid I_u^{1:T}; \Theta^{old}) + \alpha - K} \tag{22}$$

$$\widehat{D}_{k,d} = \frac{\sum_u \sum_{t=1}^T p(Z_u^t = k, D_u^t = d \mid I_u^{1:T}; \Theta^{old}) + \alpha_d - 1}{\sum_u \sum_{t=1}^T \sum_{d=1}^M p(Z_u^t = k, D_u^t = d \mid I_u^{1:T}; \Theta^{old}) + \alpha - M} \tag{23}$$

$$\hat{\theta}_k(I_j) = \frac{\sum_u \sum_{t=1}^T P(Z_u^t = k \mid I_u^{1:T}; \Theta^{old}) \sum_j 1_i(I_{uj}^t) + \alpha_j - 1}{\sum_u \sum_{t=1}^T P(Z_u^t = k \mid I_u^{1:T}; \Theta^{old}) N_u^t + \alpha - |I|} \tag{24}$$

The parameters $r$, $p$ cannot be solved in closed form. So, we employ Newton's method to get their approximate numerical solution (see Appendix C for details). The estimated parameters from formulas (21) to (24) and the numerically solved $r$ and $p$ specify the model based on the training data.

## D. *Prediction*

To make use of our proposed HSMM framework for recommendation, we need to predict the possibility that a user will consume an item. For this purpose, we need to predict each user's latent state at the next time period based on the state inferred from previous observations and calculate the probability of item consumption based on the predicted state.

When time jumps from *t-1* in the training data to *t* in the testing data, there are two possibilities: 1) user *u* stays in the last state, and 2) user *u* jumps to the next state. The probability of a state lasting from *t-d+1* to *t* is $P(Z_u^{t-d+1:t} = k)$. If $Z_u^{t-d+1:t} = k$(d=1), $Z_u^{t+1-d+1:t+1} = k$(d = 2), ..., $Z_u^{t+M-1-d+1:t+M-1} = k$(*d*=*M*), the user's latent state jumps to the next state from *t-1* to *t*. If $Z_u^{t-d+1:t} = k(1<d\le M)$, $Z_u^{t+1-d+1:t+1} = k(2 < d \le M)$, ..., $Z_u^{t+M-1-d+1:t+M-1} = k(M<d\le M)$, the user's latent state remains unchanged at *t*. Hence, the probability of item *i* being included in user *u*'s consumption at time *t* can be calculated as:

$$
\begin{aligned}
P(\mathrm{i}\in \mathrm{I}_\mathrm{u}^\mathrm{t}) &= \sum_k \sum_{d=1}^{M} \mathrm{P}(\mathrm{i}\in \mathrm{I}_\mathrm{u}^\mathrm{t} \mid Z_\mathrm{u}^{t-d+1:t} = \mathrm{k})\,\mathrm{P}(Z_\mathrm{u}^{t-d+1:t} = \mathrm{k}) + \sum_k \sum_{d=2}^{M} \mathrm{P}(\mathrm{i}\in \mathrm{I}_\mathrm{u}^\mathrm{t} \mid Z_\mathrm{u}^{t+1-d+1:t+1} = \mathrm{k})\,\mathrm{P}(Z_\mathrm{u}^{t+1-d+1:t+1} = \mathrm{k}) + \\
&\quad \ldots + \sum_k \sum_{d=M}^{M} [\mathrm{P}(\mathrm{i}\in \mathrm{I}_\mathrm{u}^\mathrm{t} \mid Z_\mathrm{u}^{t+M-d:t+\mathrm{M}-1} = \mathrm{k}) \bullet \mathrm{P}(Z_\mathrm{u}^{t+M-d:t+\mathrm{M}-1} = \mathrm{k})] \\
&= \sum_k \sum_{s\ge t}^{t+M-1} \sum_{d=s-t+1}^{M} P(\mathrm{i}\in \mathrm{I}_\mathrm{u}^\mathrm{t} \mid Z_\mathrm{u}^{s-d+1:s} = \mathrm{k})\,\mathrm{P}(Z_\mathrm{u}^{s-d+1:s} = \mathrm{k})
\end{aligned}
\tag{25}
$$

where $I_u^t$ is the items selected by user *u* at time period *t*.

The probability of a user staying in a state from *t-d+1* to *t* can be derived from the segmental state transition probability *A* and state duration probability *D* based on the last state of the training data:

$$
\begin{aligned}
&P(Z_\mathrm{u}^{t-d+1:t} = \mathrm{k}) \\
&= \sum_{i\in S\backslash\{k\}} \sum_{d'=1}^{\mathrm{M}} [P(Z_\mathrm{u}^{t-d+1:t} = \mathrm{k} \mid Z_\mathrm{u}^{t-d-d'+1:t-d} = i) P(Z_\mathrm{u}^{t-d-d'+1:t-d} = i)] \\
&= \sum_{i\in S\backslash\{k\}} \sum_{d'=1}^{M} A_{ik} D_{k,d} P(Z_\mathrm{u}^{t-d} = i, D_u^{t-d} = d')
\end{aligned}
\tag{26}
$$

$P(i \in I_u^t | Z_u^{s-d+1:s} = k)$ in formula (25) is the probability for user *u* to consume an item *i* given her predicted state *k* at *t*. According to the law of total probability, we calculate it by summing over the probability of consuming a different number of items.

$$
P(\mathrm{i}\in \mathrm{I}_\mathrm{u}^\mathrm{t} \mid Z_\mathrm{u}^{s-d+1:s} = \mathrm{k}) = \sum_{N_\mathrm{u}^\mathrm{t}=0}^{\infty} \mathrm{P}(\mathrm{i}\in \mathrm{I}_\mathrm{u}^\mathrm{t} \mid N_\mathrm{u}^\mathrm{t};\theta_k)\,\mathrm{P}(N_\mathrm{u}^\mathrm{t};\mathrm{r}_{k,d},\mathrm{p}_{k,d})
\tag{27}
$$

Since the distribution over the items follows a multinomial distribution, the probability that an item is observed in the $N_u^t$ items is:

$$\begin{aligned} &\mathrm{P}(\mathrm{i} \in \mathrm{I}_{\mathrm{u}}^{\mathrm{t}} \mid N_{\mathrm{u}}^{\mathrm{t}}; \theta_k) = 1 - P(\mathrm{i} \notin \mathrm{I}_{\mathrm{u}}^{\mathrm{t}} \mid N_{\mathrm{u}}^{\mathrm{t}}; \theta_k) \\ &= 1 - (1 - \mathrm{P}(\mathrm{i} \mid \theta_{\mathrm{k}}))^{N_{\mathrm{u}}^{\mathrm{t}}} = 1 - (1 - \theta_{\mathrm{ki}})^{N_{\mathrm{u}}^{\mathrm{t}}} \end{aligned} \tag{28}$$

Based on formula (28), considering the number of selected items follows NBD, formula (27) becomes:

$$\begin{aligned} &\mathrm{P}(\mathrm{i} \in \mathrm{I}_{\mathrm{u}}^{\mathrm{t}} \mid Z_{\mathrm{u}}^{\mathrm{t}} = \mathrm{k}) = \sum_{N_{\mathrm{u}}^{\mathrm{t}}=0}^{\infty} \mathrm{P}(\mathrm{i} \in \mathrm{I}_{\mathrm{u}}^{\mathrm{t}} \mid N_{\mathrm{u}}^{\mathrm{t}}; \theta_k) \mathrm{P}(N_{\mathrm{u}}^{\mathrm{t}}; \mathrm{r}_{k,d}, \mathrm{p}_{k,d}) \\ &= 1 - \sum_{N_{\mathrm{u}}^{\mathrm{t}}=0}^{\infty} NBD(N_{\mathrm{u}}^{\mathrm{t}}; \mathrm{r}_{k,d}, \mathrm{p}_{k,d})(1 - \theta_{\mathrm{ki}})^{N_{\mathrm{u}}^{\mathrm{t}}} \\ &= 1 - \sum_{N_{\mathrm{u}}^{\mathrm{t}}=0}^{\infty} \frac{\Gamma(N_{\mathrm{u}}^{\mathrm{t}} + r_{k,d})}{N_{\mathrm{u}}^{\mathrm{t}}! \Gamma(r_{k,d})} \mathrm{p}_{k,d}^{N_{\mathrm{u}}^{\mathrm{t}}} (1 - \mathrm{p}_{k,d})^{\mathrm{r}_{k,d}} (1 - \theta_{\mathrm{ki}})^{N_{\mathrm{u}}^{\mathrm{t}}} \\ &= 1 - \frac{(1 - \mathrm{p}_{k,d})^{r_{k,d}}}{(1 - \mathrm{p}_{k,d}(1 - \theta_{\mathrm{ki}}))^{\mathrm{r}_{k,d}}} \end{aligned} \tag{29}$$

Based on formulas (26) and (29), an item's prediction consumption in formula (25) then becomes:

$$\begin{aligned} &P(\mathrm{i} \in \mathrm{I}_{\mathrm{u}}^{\mathrm{t}}) = \sum_{k} \sum_{s \geq t} \sum_{d=s-t+1} \mathrm{P}(Z_{\mathrm{u}}^{\mathrm{t}} = \mathrm{k}, D_{\mathrm{u}}^{\mathrm{t}} = d)(1 - \frac{(1 - \mathrm{p}_{k,d})^{r_{k,d}}}{(1 - \mathrm{p}_{k,d}(1 - \theta_{\mathrm{ki}}))^{\mathrm{r}_{k,d}}}) \\ &= 1 - \sum_{k} \sum_{s \geq t} \sum_{d=s-t+1} [\frac{(1 - \mathrm{p}_{k,d})^{r_{k,d}}}{(1 - \mathrm{p}_{k,d}(1 - \theta_{\mathrm{ki}}))^{\mathrm{r}_{k,d}}} \sum_{i \in S \backslash \{k\}} \sum_{d'=1}^{\mathrm{D}} A_{ik} D_{k,d} P(Z_{\mathrm{u}}^{t-d} = i, D_u^{t-d} = d')] \end{aligned} \tag{30}$$

# IV. Evaluation Framework

## A. *Datasets*

To evaluate the effectiveness of our approach, we conduct experiments on three real-world datasets on information goods consumption, where users consume movies, webpages, and music:

1) The Netflix dataset, which comes from Netflix Contest [23] and contains over 100 million ratings of 17,770 movies by approximately 480,000 users from 1999 to 2005.
2) The Delicious dataset, which is extracted from the social bookmarking website, Delicious [49], containing 830,000 users' bookmarking of 33 million webpages from 2003 to 2007.
3) The Last.fm dataset, which is extracted from an Internet music radio website, Last.fm [50], containing

992 users' consumption of 177,000 artists' albums/songs from 2005 to 2009.

To keep the data size manageable, we filter out some less frequent users and items. Table II shows the criteria for filtering and the size of the filtered datasets.

TABLE II
FILTERING OF THE DATASETS

| | Netflix | Delicious | Last.fm |
|---|---|---|---|
| # users for each item | >2000 | >3000 | >0 |
| # items for each user | > 2000 | > 3000 | > 50 |
| After filtering | | | |
| # users | 1,212 | 1,126 | 992 |
| # items | 5,264 | 782 | 3,069 |
| # user–item interactions | 2,534,213 | 561,938 | 12,543,047 |
| Starting time | Jan 2000 | July 2004 | Aug 2005 |
| Finishing time | Dec 2005 | Dec 2007 | June 2009 |

Fig. 3 shows the evolution of number of users, items, and consumptions in each month according to the time when the users rate/bookmark/listen to the movies/webpages/music. The three numbers generally increase over time, except that the number of items in the Last.fm dataset gets stable pretty quickly.

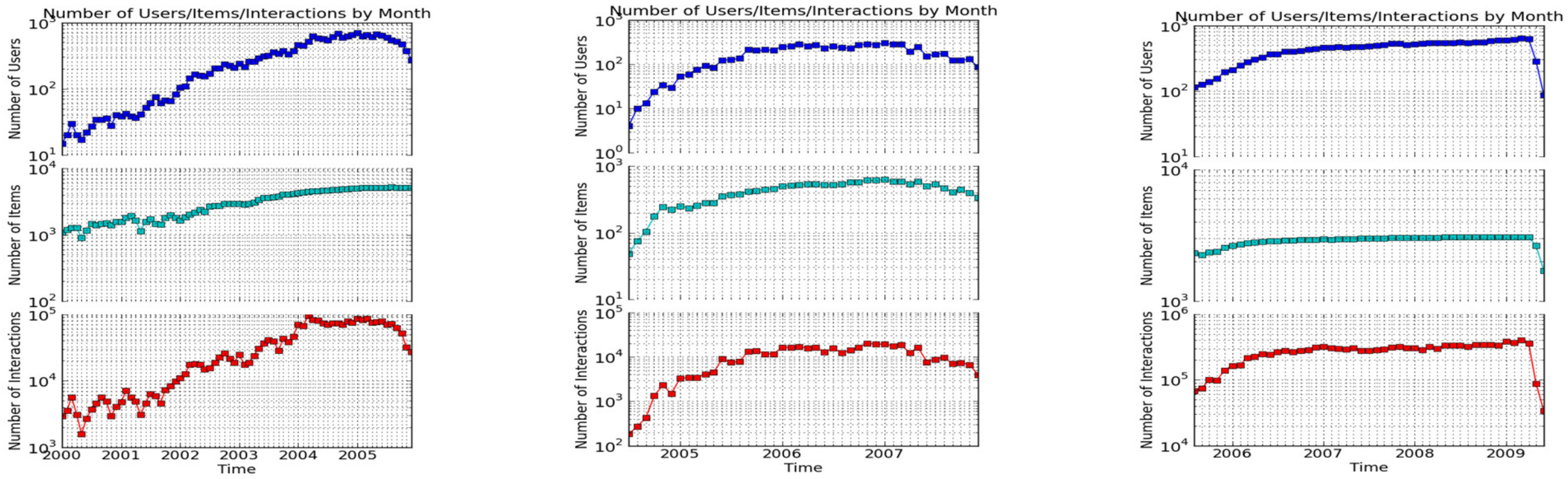

(a) Netflix (b) Delicious (c) Last.fm

Fig. 3. The number of users/items in each month of our datasets

## B. Experiment Setup

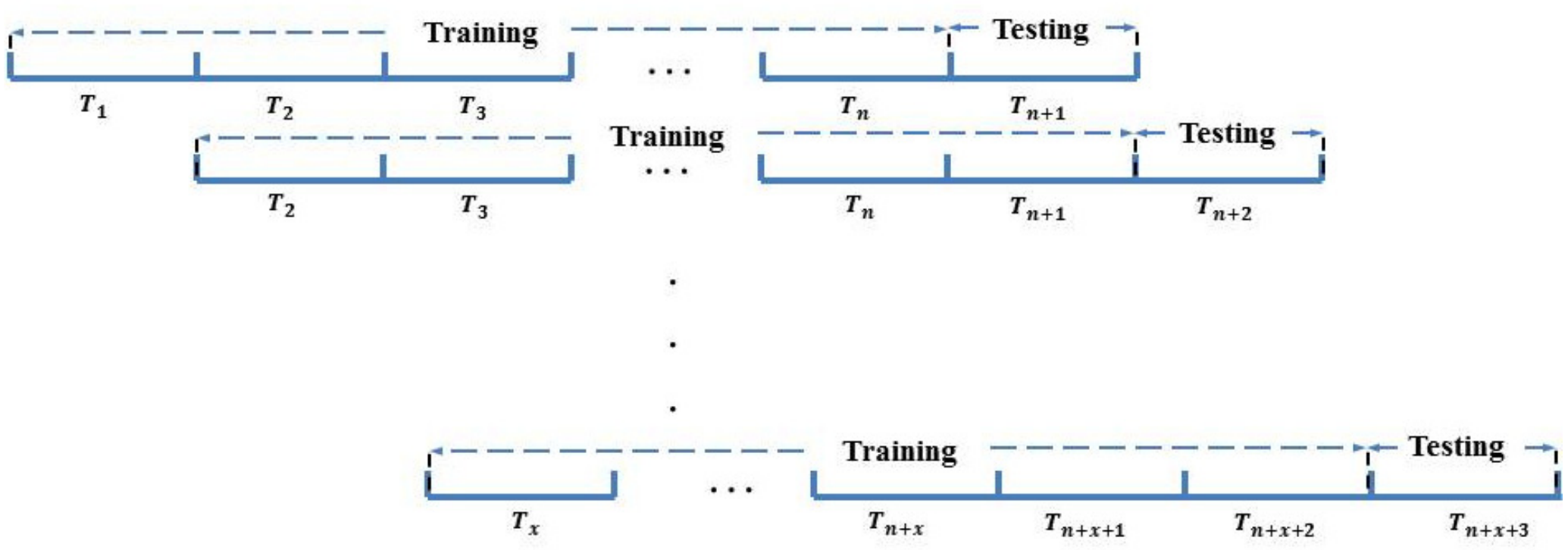

Fig. 4. The experiment rolling by month

To evaluate our model's capability to predict the temporal change of user interest, we conduct experiments on a long time-period in a rolling fashion, as shown in Fig. 4. Specifically, we consider users' state transition at the monthly level. We choose the first *n* months as training data to tune the model and make predictions on the testing data in the (*n+1*)-*th* month. Then we shift the time window of training data by one month and use the data from the 2nd month to the (*n+1*)-*th* month as training data and use the (*n+2*)-*th* month for testing[2]. We take about two-thirds of the data for training, which provides us n=48 months, 30 months, and 34 months of training data for Netflix, Delicious, and Last.fm, respectively. We continue rolling by month to the end of the dataset. Finally, we conduct 24, 12, and 13 rounds of experiments for Netflix, Delicious, and Last.fm, respectively. The evaluation metrics are computed for each train-test set for the top-5 and top-10 predictions. We repeated each experiment setup 10 times to conduct pairwise t-tests for performance comparison.

### C. *Evaluation Metrics*

In our research, we adopted the widely used precision, recall, and F-measure of the top-N recommendations for evaluation. Precision (P) is defined as the percent of correct predictions (appearing in the testing data) among all predictions. Recall (R) is the percent of correct predictions in all items existing in the test data. Because these two metrics are inversely related, we use the F–measure, F1=2×P×R/(P+R), to combine precision and recall.

### D. *Baseline Algorithms*

In this study, we chose four state-of-the-art time-dependent methods and three classic static methods as baseline algorithms. The time-dependent methods include:

1. **HMM Model** [28]. Our proposed HSMM is based on the HMM, which models transitive latent states of user interests. The model is the most similar algorithm to our approach among all baseline algorithms.

[2] Our experiment setup is slightly different from (Sahoo et al. 2012). Sahoo et al.'s (2012) training data starts from the beginning of the dataset and thus has a variant length. We consider a fixed training data length is more straightforward for performance comparison. The experiment setup causes slight performance changes, which does not affect the relative performance advantage of HSMM over our baseline algorithms.

Using it for comparison can illustrate the performance improvement caused by our design feature: the heterogeneity of user interest durations.

2. **Katz-CWT (KC) algorithm** [16]. The model employs a time-decaying method to introduce a collapsed weighted tensor into the Katz method and uses a truncated SVD for link prediction. The model's performance is among the best of existing time-dependent recommendation algorithms. It first aggregates the adjacency matrix $A_t$ with time-decay factor at each time period $t$.

$$A = \sum_{t=1}^{T}(1-\theta)^{T-t}A_t;\ \theta \in (0,1) \tag{31}$$

Then, this method conducts SVD on matrix $A$ to get the first $k$ singular values $\sigma_k$, and their corresponding singular vectors $U_k$ and $V_k$. The prediction is:

$$R = U_k \Psi_k V_k^T \tag{32}$$

Each diagonal element in $\Psi_k$ is $\Psi_k = \beta\delta_k/(1-\beta^2\ \delta_k^2)$. In the experiments, we set $\beta$ to 0.001 following [16], and tune θ from 0.1 to 0.9 in increases of 0.1.

3. **Temporal Item-based Collaborative Filtering (tIB)** [8]. The model incorporates a time-decaying function into the classic item-based collaborative filtering method. It selects nearest neighbor items based on the Pearson Correlation measurement and integrates the neighbors' ratings as:

$$r_{ui} = \frac{\sum_j r_{uj} sim(I_j, I_i) f(t_{uj})}{\sum_c sim(I_j, I_i) f(t_{uj})} \tag{33}$$

where $r_{uj}$ is the rating user $u$ gave to item $j$ at time $t_{uj}$. $\mathrm{f(t)} = \exp(-\lambda \mathrm{t})$ represents a time-decay function to reduce the weight of items over time. We tune λ from 0.1 to 0.9 in increases of 0.1.

4. **Time singular value decomposition (timeSVD)** [21]. The model incorporates time-related variables into the famous matrix factorization method. It allows for time-related user bias in matrix factorization. The predicted rating of user $u$ to item $i$, $r_{ui}$, is calculated as:

$$r_{ui} = \mu + b_t + (b_u + x_u^T z_\tau) + (b_i + s_i^T y_w) + (p_u^T q_i + \sum_k g_{uk} l_{ik} h_{\tau k}) \tag{34}$$

where μ is the general average showing the overall level of ratings, $b_t$ is time bias, $(b_u + x_u^T z_\tau)$ is user bias, $(b_i + s_i^T y_w)$ is item bias, and $(p_u^T q_i + \sum_k g_{uk} l_{ik} h_{\tau k})$ is users' preferences. We follow [21] to tune this model.

The three classic static methods include:

1. **Link Analysis (LA)** [51]. The model infers the relationship among users and items in the same fashion as the HITS algorithm. Denoting PR and CR as a product representativeness matrix and a consumer representativeness matrix, which are calculated as:

$$PR = A^T \times CR \tag{35}$$

$$CR = B \times PR + CR^0 \tag{36}$$

where each element of $B$ is $b_{ij} = a_{ij} / \left(\sum_j a_{ij}\right)^\gamma$. When the calculation of PR converges, it will be used to predict user $u$'s rating of item $i$, i.e., $R_{u,i}=PR_{i,u}$.

2. **Probabilistic latent semantic analysis (pLSA)** [52]. The model conducts probabilistic clustering on users and items to clusters $z$. The consumption prediction is calculated based on the law of total probability:

$$P(i|u;\theta) = \sum_z P(i|z)P(z|u) \tag{37}$$

3. **User-based collaborative filtering (UB)** [53]. As one of the classic recommendation methods, it uses the Pearson Correlation [54], [55] to measure the similarity between users and make cross-recommendations.

## V. Results

TABLE III
PERFORMANCE COMPARISON ON NETFLIX DATASET#

| | Top-5 Recommendation | | | Top-10 Recommendation | | |
|---|---|---|---|---|---|---|
| Alg. | P | R | F1 | P | R | F1 |
| HSMM | **0.1236**** | **0.0244*** | **0.0405**** | **0.1101**** | **0.0428**** | **0.0606**** |
| HMM | 0.0757 | 0.0151 | 0.0248 | 0.0677 | 0.0268 | 0.0377 |
| KC | 0.1052 | 0.0216 | 0.0353 | 0.0920 | 0.0368 | 0.0516 |
| tIB | 0.0346 | 0.0068 | 0.0112 | 0.0323 | 0.0127 | 0.0179 |
| timeSVD | 0.0150 | 0.0031 | 0.0051 | 0.0121 | 0.0049 | 0.0069 |
| LA | 0.0256 | 0.0050 | 0.0082 | 0.0238 | 0.0092 | 0.0130 |
| pLSA | 0.0154 | 0.0031 | 0.0050 | 0.0156 | 0.0062 | 0.0086 |
| UB | 0.0262 | 0.0051 | 0.0084 | 0.0242 | 0.0095 | 0.0133 |

# HSMM: K=40,M=5; HMM: K=40; Katz-CWT: θ=0.8; tIB: λ=0.9; timeSVD: K=120; pLSA: K=30. p<0.1 *;
The significance level between the largest and the second largest value in each column: p<0.1 *; p<0.05 **; p< 0.001 ***;

TABLE IV

PERFORMANCE COMPARISON ON DELICIOUS DATASET#

| | Top-5 Recommendation | | | Top-10 Recommendation | | |
|---|---|---|---|---|---|---|
| Alg | P | R | F1 | P | R | F1 |
| HSMM | **0.0096****| **0. 0406**** | **0.0153**** | **0.0083**** | **0.0603*** | **0.0145**** |
| HMM | 0.0050 | 0.0184 | 0.0077 | 0.0048 | 0.0345 | 0.0083 |
| KC | 0.0082 | 0.0309 | 0.0128 | 0.0076 | 0.0564 | 0.0133 |
| tIB | 0.0047 | 0.0181 | 0.0074 | 0.0041 | 0.0290 | 0.0072 |
| timeSVD | 0.0029 | 0.0096 | 0.0044 | 0.0033 | 0.0229 | 0.0058 |
| LA | 0.0028 | 0.0101 | 0.0043 | 0.0034 | 0.0245 | 0.0059 |
| pLSA | 0.0038 | 0.0134 | 0.0058 | 0.0040 | 0.0269 | 0.0070 |
| UB | 0.0035 | 0.0134 | 0.0055 | 0.0029 | 0.0219 | 0.0051 |

[#] HSMM: K=30,M=5; HMM: K=30; Katz-CWT: θ=0.4; tIB: λ=0.6; timeSVD: K=110; pLSA: K=40.
The significance level between the largest and the second largest value in each column: p<0.1 *; p<0.05 **; p< 0.001 ***;

TABLE V
PERFORMANCE COMPARISON ON LAST.FM DATASET [#]

| | Top-5 Recommendation | | | Top-10 Recommendation | | |
|---|---|---|---|---|---|---|
| Alg | P | R | F1 | P | R | F1 |
| HSMM | **0.0107** | **0.0109** | **0.0101** | **0.0099** | **0.0204** | **0.0126** |
| HMM | 0.0085 | 0.0090 | 0.0081 | 0.0076 | 0.0166 | 0.00971 |
| KC | **0.0102** | **0.0102** | **0.0095** | **0.0092** | **0.0191** | **0.0118** |
| tIB | **0.0102** | **0.0102** | **0.0096** | **0.0103** | **0.0201** | **0.0131** |
| timeSVD | 0.0039 | 0.0040 | 0.0038 | 0.0039 | 0.0074 | 0.0049 |
| LA | 0.0086 | 0.0089 | 0.0083 | 0.0081 | 0.0159 | 0.0104 |
| pLSA | **0.0104** | **0.0094** | **0.0096** | 0.0089 | 0.0173 | 0.0113 |
| UB | **0.0102** | **0.0100** | **0.0096** | **0.0104** | **0.0200** | **0.0133** |

[#] HSMM: K=40,M=4; HMM: K=30; Katz-CWT: θ=0.9; tIB: λ=0.1; timeSVD: K=60; pLSA: K=50.
The bold numbers are not significantly different from the largest number in each column.

Tables III to V report the different algorithms' performances on the three real-world datasets. The parameters of the algorithms are tuned to their best performances, which are also reported in the tables. In the tables we highlight the values that are not significantly different from the highest value in each column and determine the significance level by comparing the highest value vs. the second highest value. As we can see, the performance of our proposed approach is always in the group of best models. On the Netflix and Delicious datasets, our proposed approach significantly outperforms all baseline methods (at 90% to 95% confidence level) and achieves about 10%~ 20% performance improvements over the second best model. In the following sections, we further discuss the models' performances and possible reasons for their performances.

## A. *HSMM vs. HMM and User Interest Heterogeneity*

TABLE VI
COMPARISON OF HSMM AND HMM PERFORMANCE

| p-value | Top-5 Recommendation | | | Top-10 Recommendation | | |
|---|---|---|---|---|---|---|
| | P | R | F1 | P | R | F1 |
| Netflix | 0.000 | 0.000 | 0.000 | 0.000 | 0.000 | 0.000 |
| Delicious | 0.000 | 0.000 | 0.000 | 0.000 | 0.000 | 0.000 |
| Last.FM | 0.000 | 0.000 | 0.000 | 0.000 | 0.000 | 0.000 |

Our HSMM was rooted in the HMM. Table VI shows the results of pairwise t-tests on the two models. As we can see, in all datasets the HSMM method consistently and significantly (at 99.9% confidence level) performs better. As shown in Tables III to V, the relative performance improvement of HSMM over HMM is 15%~55%.

The major difference between HSMM and HMM is that HSMM captures the heterogeneity of users' interest durations. In HMM, the state self-transition allows user interest duration to follow a geometric distribution, where a longer duration has a smaller probability. In HSMM, we take a non-parametric approach and allow state duration to follow any shape of distributions, which is learned from the data. This design allows HSMM to better fit the evolution of users' interests in different contexts and applications. Furthermore, as shown in formula (25), the prediction of our proposed HSMM makes use of multiple time periods' data related to state $Z_u^{s-d+1:s}$. Such a design feature may also cause performance improvements as compared with the HMM, which only makes use of the last time period for prediction.

To further illustrate the ability of the HSMM to capture user interest duration heterogeneity, we conduct further analysis of the state durations generated by the HSMM and HMM models. Particularly, we inspect the distribution of state duration for users. As shown in Tables III to V, the HSMM has a maximum duration $M$ equal to 4 or 5 after tuning the parameters. So, there are only six possible types of duration distributions as shown in Fig. 5-I: (a) an inverse U shape distribution with more medium length durations; (b) a U shape distribution with more short or long length durations; (c) a negative exponential distribution, where longer duration has smaller probability; (d) an exponential duration, where shorter duration has smaller probability; (e) a uniform distribution, where long and short durations have the same probability; (f) a N shape or inverse N shape distribution, where the relationship between interest duration length and probability is complicated.

We aggregate users according to the six types of state duration distributions. Fig. 5-II presents the portion of users with different state duration distributions as identified by HSMM and HMM. Obviously, under the HMM, most users are considered to have a smaller (or similar) probability to stay in a state longer. However, our HSMM can identify a significant portion of users having other shapes of state duration dis-tributions. For example, in the Last.FM dataset, we can observe about 37% of users following a U shape distribution, i.e., having many periods of rapidly switching interests and many periods of long-term interests. Clearly, HSMM is better than HMM in identifying interest state duration heterogeneity.

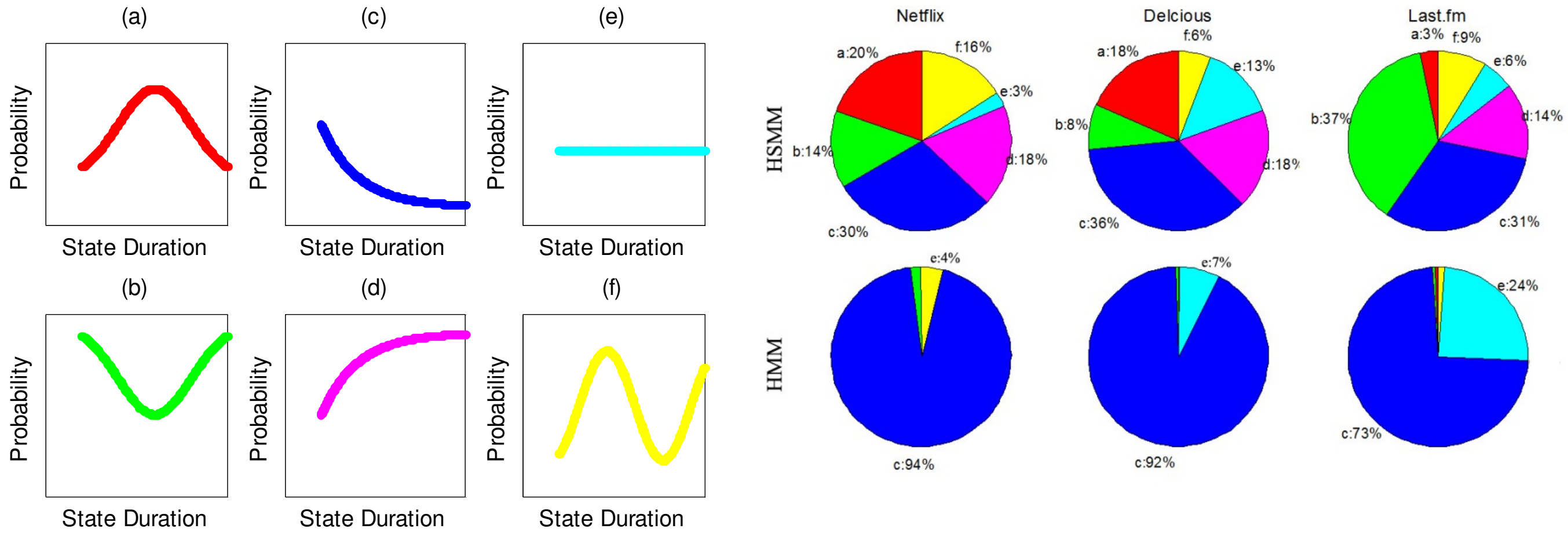

I) Possible Shapes of Interest State Duration Distribution

II) The Portion of Users with Different State Duration Distributions

Fig. 5 User interest state duration heterogeneity identified by HSMM and HMM

### *B. Comparison with Other Time-dependent Methods and Static Methods*

In addition to HMM, we applied three other time-dependent methods, Katz-CWT, tIB, and timeSVD. Both Katz-CWT and tIB model users' interest changes with the time-decay factor. The HSMM method has better performance than these two methods on both the Netflix and Delicious datasets. On the Last.fm dataset, these methods have similar performances, which will be discussed in subsection V-D. We believe the advantage of the HSMM method mainly comes from the flexible assessment of state transition of users. Katz-CWT and tIB methods assume a constant decaying tendency across users and across different time periods. For some users, that will lose information due to the underweighting of old data [23]. The HSMM allows different changing tendencies across different states (i.e., state duration heterogeneity) and allows different users in different states (i.e., user interest heterogeneity, which is also captured by HMM). So, some users can stay in some states for a longer time, and the information of historical data is reflected in latent states. It avoids the difficulty of carrying on historical information and identifying changes of information in one time-decay factor. Thus, we observe HSMM's advantage over time-decay methods as we predicted.

The timeSVD method failed to a large extent. We believe this is because timeSVD was designed for (numerical) rating prediction, rather than (binary) consumption relationship prediction. A similar phenomenon on the matrix factorization methods' performance was also observed in previous studies [56].

From our experiments, we can also observe the advantage of time-dependent methods over static methods in the Netflix and Delicious datasets (except for timeSVD, which is due to the unfitting to the problem setup). This illustrates that the changes of users' interests do exist in the datasets. In Last.fm, the two types of methods do not have a systematic difference, which may be due to relatively stable user interests in music consumption.

### C. *Sensitivity Analysis*

The HSMM model has two parameters that need to be specified by the modeler, the number of latent states (*K*) and the allowed maximum length of state duration (*M*). We conduct additional experiments to assess the sensitivity of our model to these two parameters.

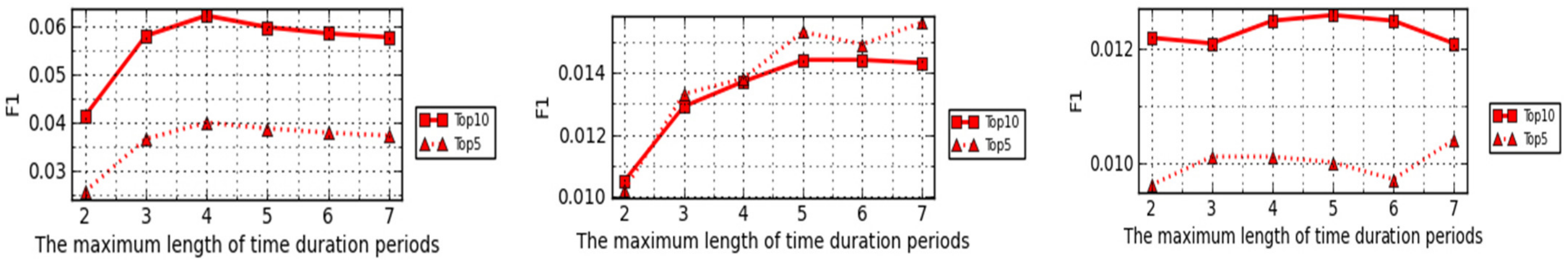

(a) Netflix (K=40) (b) Delicious (K=30) (c) Last.fm (K=40)

Fig. 6. HSMM performance with different maximum length of state duration

Fig. 6 reports the change of performance with increasing values of *M* while keeping *K* constant (as the value that can provide the best performance in their dataset). Here we only report the F-measure since the curves on precision and recall are similar. The prediction performances generally show an increasing trend, which stabilizes when *M* reaches 4 or 5, in both top-10 and top-5 recommendations. The performance change is relatively less in the Last.fm dataset. When $M=1$, the HSMM and HMM are essentially the same. When *M* is really small, the ability for HSMM to capture users' long-term interests is limited, which leads to the relatively low performance. The result further illustrates the importance of capturing the heterogeneity of users' interests with longer duration in time-dependent recommendation.

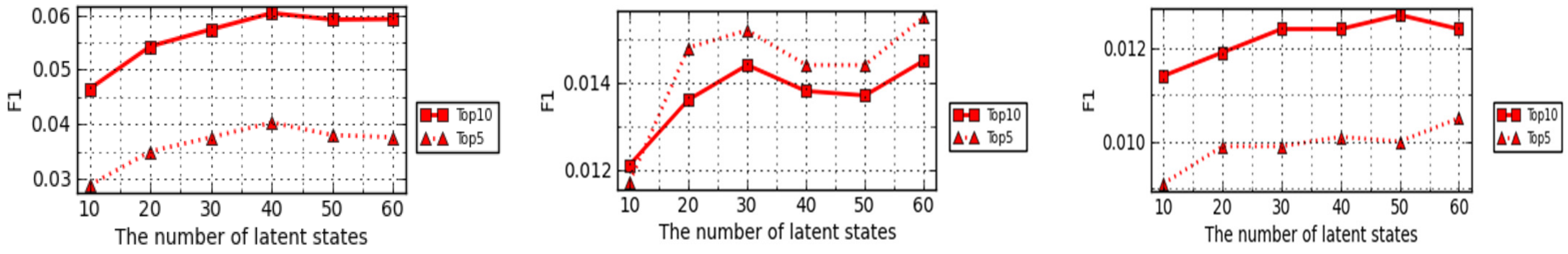

(a) Netflix (M=5) (b) Delicious (M=5) (c) Last.fm (M=4)

Fig. 7. HSMM performance with different number of latent states

Fig. 7 reports the results of adjusting the number of latent states (*K*) in HSMM. Similar to Fig. 6, we only show the F-measure curves with *M* set to its optimal value. As we can see, HSMM's performance increases with *K* and stabilizes when *K* becomes large. Increasing the number of hidden states increases the complexity of the model and its capability to capture the heterogeneity of user interests. The value of modeling user heterogeneity was already shown in the HMM. Our HSMM approach further illustrates the importance of this design feature.

### D. *Post hoc Analysis: The Impact of Data Characteristics*

As one may notice, in our experiments the HSMM (and the HMM) shows less advantage over other methods in the Last.FM dataset as compared with other two datasets. So, what characteristics of the Last.FM data may affect the performance of the two transitive state methods?

First, both models employ latent states to capture user interest heterogeneities. In Fig. 8, we inspect the portion of users according to the number of states they have. We found that the users in Netflix and Delicious generally vary across a larger number of states as compared with the users in the Last.fm dataset. In Netflix and Delicious, users may switch between (on average) 9 and 14 states. But most Last.fm users only have 3~5 states. Clearly, the Netflix and Delicious datasets have a higher level of interest heterogeneity over time. In fact, users are more likely to listen to songs created by the artists they preferred before. People show less interest change in listening to music as compared with watching movies or reading webpages. With a small number of interest states, the benefit of modeling the switch of latent interest states (in HSMM and HMM) would be smaller.

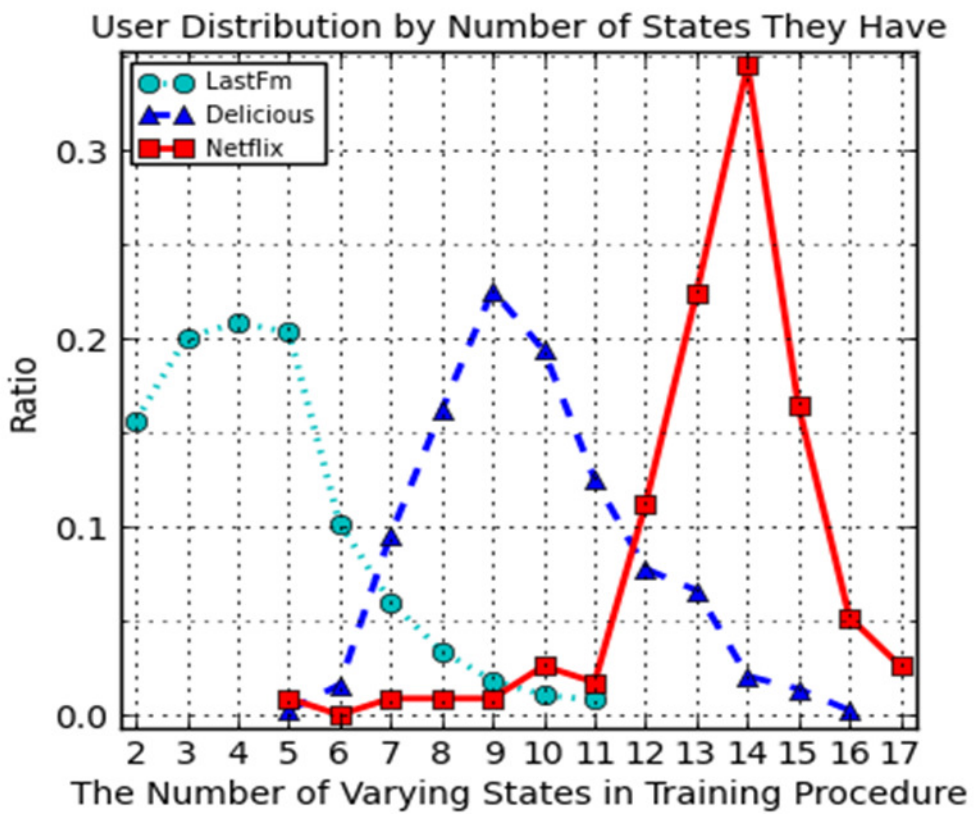

Fig. 8. User distribution by number of latent states (K=40, M=5)

Then we inspect 5 states that were "visited" by the largest number of users. Fig. 9 visualizes the numbers of users in such 5 states of each dataset in each month. As we can see, in Netflix and Delicious there are clear patterns of the rise and fall of users' interests in the five states. Apparently, such user behaviors are affected by events, such as the release of new movies or appealing news events. However, in Last.fm, the number of users in the top-5 states is much more stable, showing the stable nature of music consumption. Since users' collective interest does not show strong transitions in Last.fm, it is difficult to show the advantage of HSMM in capturing changing interests of users for time-dependent recommendation.

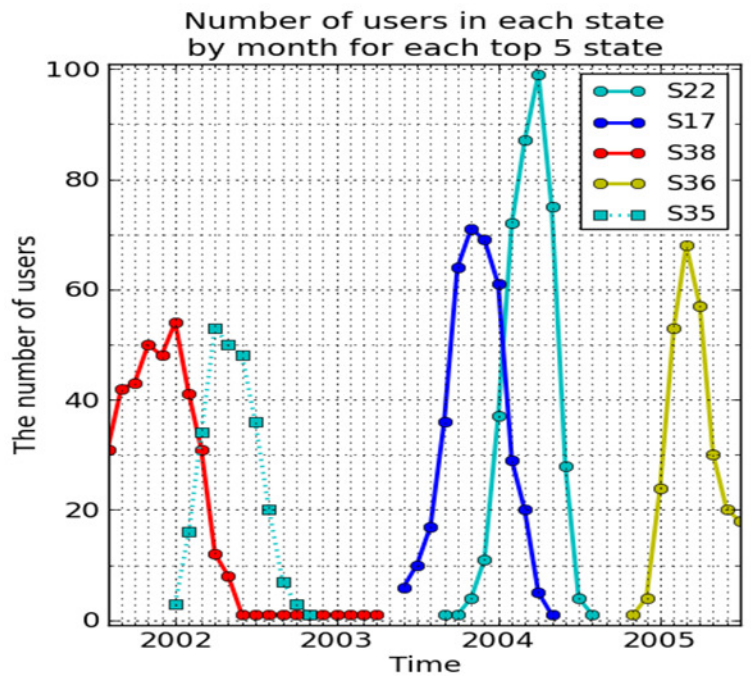

(a) Netflix (K=40, M=5)

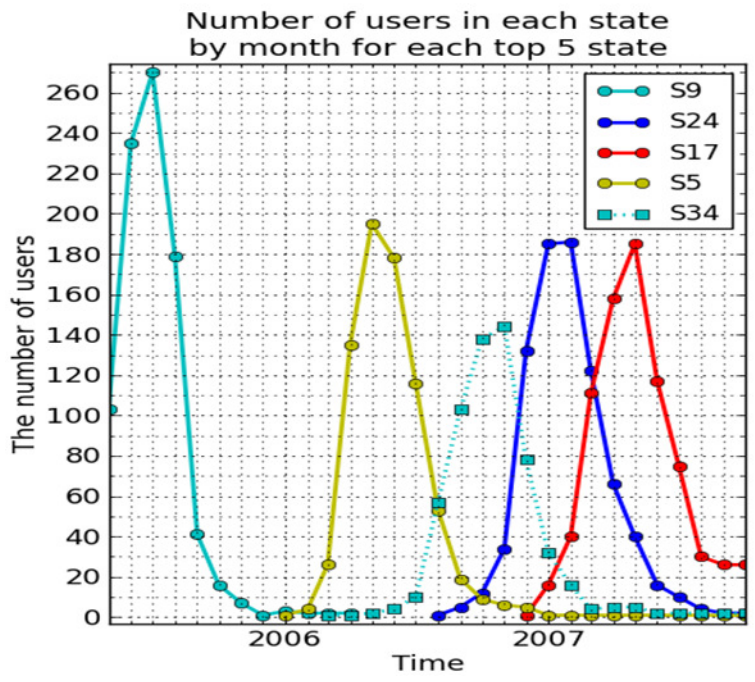

(b) Delicious (K=40, M=5)

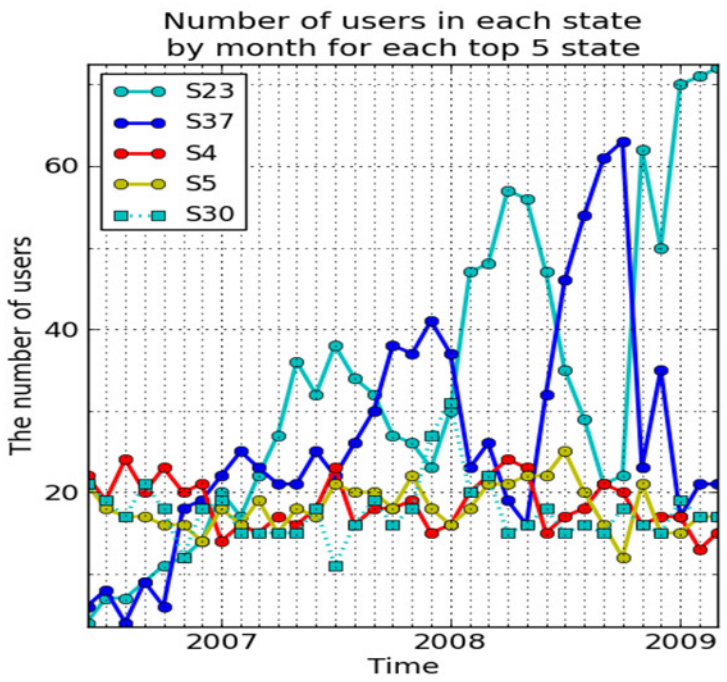

(c) Last.fm (K=40, M=5)

Fig. 9. Number of users in each state by month for top-5 states

### *E. Discussion, Limitations, and Implications*

The experiment results clearly demonstrate the advantage of our proposed HSMM approach over existing methods with a large performance improvement. On datasets where users have clear changes of interests, more changes of interests (i.e., more switches across latent states), and with heterogeneous length of interest durations, our proposed HSMM approach shows a clear advantage over existing methods. We also find that the HSMM performance first increases and then stabilizes when allowing for more latent states and longer maximum state durations.

The performance improvement of HSMM may come from the modeling of heterogeneous user interests and interest durations. Our proposed HSMM method inherits the advantage of the HMM model in employing latent states to model different interests of users, i.e., the heterogeneity of users' interests, and using the transition of states to track users' drifting interests. Furthermore, the HSMM model takes a non-parametric approach to model state duration distribution, which allows capturing different duration lengths of users' interests, i.e., the heterogeneity of state duration. Combining the two design features empowers the HSMM to further illustrate the value of transitive state methods by capturing not only the rapid changing interests but also the long-term (but still changing) interests. It significantly outperforms its ancestor, the HMM, which assumes shorter durations with larger probabilities (i.e., better modeling rapidly changing interests).

In addition to user interest heterogeneity and user interest duration heterogeneity, there may exist other types of heterogeneity among users, such as state transition heterogeneity. In practice, the experience in the last few periods may affect the transition of users' interests. For example, users may have a different probability of switching from R&B music to rock music if they experience each differently during daily work. The transitive probability may change over time, i.e., the Markov chain may have a non-stationary state transition. Our model follows the classic HSMM setup and does not consider such a situation. However, there exist studies on modeling non-stationarity state transition. For example, Lanchantin and Pieczynski [57] and Lanchantin*, et al.* [58] introduced a second random chain to control the non-stationary transition process in HMM. Lapuyade-Lahorgue and Pieczynski [32] extended the HSMM by defining a minimum state duration and allowing states' self-transition. These prior studies enrich the Markov modeling in terms of the complicated transition process and show the possibility of modeling state transition heterogeneity in time-dependent recommendation. We believe such extensions have the potential to further improve recommendation performance in the evolution of users' interests.

Due to the complexity of the model, our model takes longer than HMM for parameter tuning. However, the prediction setup of the HSMM is fast and can be done in real time. After offline parameter tuning, the HSMM can support online prediction in real-world applications.

Our proposed approach in time-dependent recommendation has significant implications for practice, especially in the knowledge management domain. As compared with e-commerce applications, knowledge management applications often have more time-related factors. For example, in an education context, the recommendation of educational materials should be aligned with the school calendar, which is the nature of the curriculum design. In a library or document recommendation context, people's interest may be affected by the trend of readership (such as the following of Nobel Prize winners' books each year) and also the development of inherent interests when users grow. In such applications, considering the time feature and interest durations in recommendation will bring even more benefits.

## VI. Conclusion and Future Directions

In our research, we develop an HSMM-based approach to capture the changes of users' interests. This model allows each latent state to last a different amount of time, which can model the heterogeneous duration of users' interests. We derive an EM algorithm to estimate the parameters of the model and make predictions based on the transition and duration of states. We compare the model with the state-of-the-art time-dependent and classic static methods on three real-world datasets. The performance of our method is consistently good and improves significantly over the benchmark algorithms, especially on datasets with more user interest transitions. The experiment results also show the existence of heterogeneity of state durations in the datasets.

In the future, we will continue to study time-dependent recommendation models. We would like to model non-stationary transition, such as Lapuyade-Lahorgue and Pieczynski [32], for the purpose of time-dependent recommendation. We will also explore different setups to model state duration, such as re-laxing the bound on state duration. Second, we will study performance of HSMM/HMM under different granularities of time periods and examine the modeling of within-state heterogeneity to deal with the sparsity problem when the time period is set at a shorter level. We will also scale the model to support processing even larger datasets. Our ultimate goal is to develop a comprehensive and efficient framework that can better tackle the time-dependent recommendation problem.

# APPENDIX A

## Derivation of formulas for forward variable, backward variable, and the posterior distribution

1. The forward variable $\alpha(Z_u^t = i, D_u^t = d)$ represents the conditional probability of the value of $Z_u^{t-d+1:t} = i$ in terms of the given data up to time period $t$. It can be represented as:

$$\begin{aligned}
&\alpha(Z_u^t = \mathrm{i}, D_u^t = \mathrm{d}) = P(Z_u^{t-d+1:t} = \mathrm{i} \mid I_u^{1:\mathrm{t}}) \\
&= \sum_{j \in \mathrm{S}\backslash\{\mathrm{i}\}} \sum_{d'} P(Z_u^{t-d-d'+1} = j, Z_u^{t-d+1:t} = \mathrm{i} \mid I_u^{1:\mathrm{t}}) \\
&= \sum_{j \in \mathrm{S}\backslash\{\mathrm{i}\}} \sum_{d'} [P(I_u^{1:\mathrm{t}} \mid Z_u^{t-d-d'+1} = j, Z_u^{t-d+1:t} = \mathrm{i})\mathrm{P}(Z_u^{t-d-d'+1} = j, Z_u^{t-d+1:t} = \mathrm{i})] / P(I_u^{1:\mathrm{t}})
\end{aligned} \qquad \text{(A1)}$$

Considering the conditional independence property:

$$P(I_u^{1:\mathrm{t}} \mid Z_u^{t-d-d'+1} = j, Z_u^{t-d+1:t} = \mathrm{i}) = P(I_u^{1:\mathrm{t-d}} \mid Z_u^{t-d-d'+1} = j) P(I_u^{t-d+1:\mathrm{t}} \mid Z_u^{t-d+1:t} = \mathrm{i}) \qquad \text{(A2)}$$

Formula (A1) can be derived as:

$$\alpha(Z_u^t = i, D_u^t = d) = \sum_{j\in S\backslash\{i\}} \sum_{d'} [P(I_u^{1:t-d} \mid Z_u^{t-d-d'+1} = j) P(I_u^{t-d+1:t} \mid Z_u^{t-d+1:t} = i) P(Z_u^{t-d-d'+1} = j, Z_u^{t-d+1:t} = i)] / P(I_u^{1:t}) \quad (A3)$$

Due to:

$$P(I_u^{1:t-d} \mid Z_u^{t-d-d'+1} = j) = \frac{P(Z_u^{t-d-d'+1} = j \mid I_u^{1:t-d}) P(I_u^{1:t-d})}{P(Z_u^{t-d-d'+1} = j)} \quad (A4)$$

$$P(Z_u^{t-d-d'+1} = j, Z_u^{t-d+1:t} = i) = P(Z_u^{t-d+1:t} = i \mid Z_u^{t-d-d'+1} = j) P(Z_u^{t-d-d'+1} = j) \quad (A5)$$

The forward variable is derived as:

$$\alpha(Z_u^t = i, D_u^t = d) = \sum_{j\in S\backslash\{i\}} \sum_{d'} [\alpha(Z_u^{t-d} = j, D_u^{t-d} = d') A_{(j,d')(i,d)} P(I_u^{t-d+1:t} \mid Z_u^{t-d+1:t} = i)] / P(I_u^{t-d+1:t} | I_u^{1:t-d}) \quad (A6)$$

In our model, we define the transition probability is independent of the duration of the previous state and the duration is only dependent on the current state, i.e., $A_{(j,d')(i,d)} = P(Z_u^t = i, D_u^t = d | Z_u^{t-d} = j) = A_{ji} D_{i,d}$. Hence, we can get the last recursive forward variable:

$$\alpha(Z_u^t = i, D_u^t = d) = [\sum_{j\in S\backslash\{i\}} \sum_{d'} \alpha(Z_u^{t-d} = j, D_u^{t-d} = d') A_{ji}] P(I_u^{t-d+1:t} \mid Z_u^{t-d+1:t} = i) D_{i,d} / P(I_u^{t-d+1:t} | I_u^{1:t-d}) \quad (A7)$$

2. The backward variable $\beta(Z_u^t = i, D_u^t = d)$ is defined as the ratio of two conditional probabilities:

$$\begin{aligned}
\beta(Z_u^t = i, D_u^t = d) &= P(I_u^{t+1:T} \mid Z_u^{t-d+1:t} = i) / P(I_u^{t+1:T} \mid I_u^{1:t}) \\
&= \sum_{j\in S\backslash\{i\}} \sum_{d'} P(I_u^{t+1:T}, Z_u^{t+1:t+d'} = j \mid Z_u^{t-d+1:t} = i) / P(I_u^{t+1:T} \mid I_u^{1:t}) \\
&= \sum_{j\in S\backslash\{i\}} \sum_{d'} [P(I_u^{t+1:T} \mid Z_u^{t+1:t+d'} = j, Z_u^{t-d+1:t} = i) P(Z_u^{t+1:t+d'} = j | Z_u^{t-d+1:t} = i)] / P(I_u^{t+1:T} \mid I_u^{1:t}) \\
&= \sum_{j\in S\backslash\{i\}} \sum_{d'} P(I_u^{t+1:T}, Z_u^{t+1:t+d'} = j \mid Z_u^{t-d+1:t} = i) / P(I_u^{t+1:T} \mid I_u^{1:t})
\end{aligned} \quad (A8)$$

Considering the independent conditional probabilities:

$$P(I_u^{t+1:T} \mid Z_u^{t+1:t+d'} = j, Z_u^{t-d+1:t} = i) = P(I_u^{t+1:T} \mid Z_u^{t+1:t+d'} = j) \quad (A9)$$

$$\begin{aligned}
P(I_u^{t+1:T} \mid Z_u^{t+1:t+d'} = j) &= P(I_u^{t+1:t+d'}, I_u^{t+d'+1:T} \mid Z_u^{t+1:t+d'} = j) \\
&= P(I_u^{t+1:t+d'} \mid Z_u^{t+1:t+d'} = j) P(I_u^{t+d'+1:T} \mid Z_u^{t+1:t+d'} = j)
\end{aligned} \quad (A10)$$

Formula (A8) can be derived as:

$$
\begin{aligned}
&\beta(Z_u^t = \mathrm{i}, D_u^t = \mathrm{d}) \\
&= \sum_{j\in S\backslash\{i\}} \sum_{d'} [\mathrm{P}(I_u^{t+1:t+d'}, I_u^{t+d'+1:T} \mid Z_u^{t+1:t+d'} = \mathrm{j})\mathrm{P}(Z_u^{t+1:t+d'} = \mathrm{j} \mid Z_u^{t-d+1:t} = \mathrm{i})] / P(I_u^{t+1:\mathrm{T}} \mid I_u^{1:\mathrm{t}}) \\
&= \sum_{j\in S\backslash\{i\}} \sum_{d'} [\mathrm{P}(I_u^{t+d'+1:T} \mid Z_u^{t+1:t+d'} = \mathrm{j}) / P(I_u^{t+d'+1:\mathrm{T}} \mid I_u^{1:\mathrm{t+d'}})\mathrm{P}(I_u^{t+1:t+d'} \mid Z_u^{t+1:t+d'} = \mathrm{j})\mathrm{P}(Z_u^{t+1:t+d'} = \mathrm{j} \mid Z_u^{t-d+1:t} = \mathrm{i})] \\
&\quad / [P(I_u^{t+1:\mathrm{T}} \mid I_u^{1:\mathrm{t}}) / P(I_u^{t+d'+1:\mathrm{T}} \mid I_u^{1:\mathrm{t+d'}})] \\
&= \sum_{j\in S\backslash\{i\}} \sum_{d'} [\beta(Z_u^{t+d'} = j, D_u^{t+d'} = \mathrm{d'})\mathrm{P}(I_u^{t+1:t+d'} \mid Z_u^{t+1:t+d'} = \mathrm{j}) a_{ij} p_j(d')] / P(I_u^{t+1:\mathrm{t+d'}} \mid I_u^{1:\mathrm{t}})
\end{aligned} \tag{A11}
$$

3. $\mathrm{P}(Z_u^{t+1:t+d} = i, Z_u^{t-d'+1:t} = j \mid I_u^{1:T})$, the posterior probability that a user's latent state j with duration *d'* turns to the state j with duration d at *t* in terms of the observations can be represented as:

$$
\begin{aligned}
&P(Z_u^{t+1:t+d} = i, Z_u^{t-d'+1:t} = j \mid I_u^{1:T}) \\
&= [P(I_u^{1:T} \mid Z_u^{t+1:t+d} = i, Z_u^{t-d'+1:t} = j) P(Z_u^{t+1:t+d} = i, Z_u^{t-d'+1:t} = j)] / P(I_u^{1:T})
\end{aligned} \tag{A12}
$$

Considering the conditional dependent properties (A2), (A9), and (A10), we obtain:

$$
\begin{aligned}
&P(Z_u^{t+1:t+d} = i, Z_u^{t-d'+1:t} = j \mid I_u^{1:T}) \\
&= [P(I_u^{1:t}, I_u^{t+1:t+d}, I_u^{t+d+1:T} \mid Z_u^{t+1:t+d} = i, Z_u^{t-d'+1:t} = j) P(Z_u^{t+1:t+d} = i, Z_u^{t-d'+1:t} = j)] / P(I_u^{1:T}) \\
&= P(I_u^{1:t} \mid Z_u^{t-d'+1:t} = j) P(I_u^{t+1:t+d} \mid Z_u^{t+1:t+d} = i) P(I_u^{t+d+1:T} \mid Z_u^{t+1:t+d} = i) P(Z_u^{t+1:t+d} = i, Z_u^{t-d'+1:t} = j) / P(I_u^{1:T}) \\
&= P(Z_u^{t-d'+1:t} = j \mid I_u^{1:t}) P(I_u^{t+1:t+d} \mid Z_u^{t+1:t+d} = i) P(Z_u^{t+1:t+d} = i \mid Z_u^{t-d'+1:t} = j) [P(I_u^{t+d+1:T} \mid Z_u^{t+1:t+d} = i) / P(I_u^{t+d+1:T} \mid I_u^{1:t+d})] \\
&\quad / [P(I_u^{1:T}) / P(I_u^{1:t}) / P(I_u^{t+d+1:T} \mid I_u^{1:t+d})] \\
&= \alpha(Z_u^t = j, D_u^t = d') P(I_u^{t+1:t+d} \mid Z_u^{t+1:t+d} = i) P(Z_u^{t+1:t+d} = i \mid Z_u^{t-d'+1:t} = j) \beta(Z_u^{t+d} = i, D_u^{t+d} = d) / P(I_u^{t+1:t+d} \mid I_u^{1:t})
\end{aligned} \tag{A13}
$$

APPENDIX B

**Maximum a posteriori (MAP) estimation for the parameters**

The initial distribution $\pi$ follows a multinomial distribution, and its conjugate prior distribution is a Dirichlet distribution.

$$\pi \sim Direchlet(x \mid \alpha_1, \ldots, \alpha_K), \text{ where } \alpha_i = \alpha / K \tag{B1}$$

Hence, we can get the probability density function given by:

$$g(\pi_k) = \frac{1}{B(\alpha)} \prod_k \pi_k^{\alpha_k - 1} \tag{B2}$$

where $B(\alpha)$ is a gamma function. Hence, we have:

$$\widehat{\pi_{\mathrm{k}}} = \arg\max_{\pi_{\mathrm{k}}}\{\sum_{u}\sum_{k} P(\mathrm{Z}_{\mathrm{u}}^{1} = \mathrm{k} \mid \mathrm{I}_{\mathrm{u}}^{1:\mathrm{T}}; \Theta^{old}) \log \pi_{\mathrm{k}} + \log g(\pi_{\mathrm{k}})\} \quad \text{(B3)}$$

In order to derive this argmax function, we employ a Lagrange multiplier λ to combine the function and condition:

$$L(\pi_{\mathrm{k}}, \lambda) = \sum_{u}\sum_{k} P(\mathrm{Z}_{\mathrm{u}}^{1} = \mathrm{k} \mid \mathrm{I}_{\mathrm{u}}^{1:\mathrm{T}}; \Theta^{old}) \log \pi_{\mathrm{k}} + \sum_{k} (\alpha_{\mathrm{k}} - 1) \log \pi_{\mathrm{k}} - \log B(\alpha) - \lambda(\sum_{k} \pi_{\mathrm{k}} - 1) \quad \text{(B4)}$$

Take derivatives on $\pi$ and set it equal to 0.

$$\frac{\partial L}{\partial \pi_{\mathrm{k}}} = \sum_{u} P(\mathrm{Z}_{\mathrm{u}}^{1} = \mathrm{k} \mid \mathrm{I}_{\mathrm{u}}^{1:\mathrm{T}}; \Theta^{old}) * \frac{1}{\pi_{\mathrm{k}}} + \frac{\alpha_{\mathrm{k}} - 1}{\pi_{\mathrm{k}}} - \lambda = 0 \quad \text{(B5)}$$

which generates:

$$\widehat{\pi_{\mathrm{k}}} = [\sum P(\mathrm{Z}_{\mathrm{u}}^{1} = \mathrm{k} \mid \mathrm{I}_{\mathrm{u}}^{1:\mathrm{T}}; \Theta^{old}) + \alpha_{\mathrm{k}} - 1] / \lambda \quad \text{(B6)}$$

Due to $\sum_{\mathrm{k}} \widehat{\pi}_{\mathrm{k}} = 1$, we can get:

$$\begin{aligned} \lambda &= \sum_{k}(\sum_{u} P(\mathrm{Z}_{\mathrm{u}}^{1} = \mathrm{k} \mid \mathrm{I}_{\mathrm{u}}^{1:\mathrm{T}}; \Theta^{old}) + \alpha_{\mathrm{k}} - 1) \\ &= \sum_{u}\sum_{k} P(\mathrm{Z}_{\mathrm{u}}^{1} = \mathrm{k} \mid \mathrm{I}_{\mathrm{u}}^{1:\mathrm{T}}; \Theta^{old}) + \alpha - K \end{aligned} \quad \text{(B7)}$$

Thus:

$$\widehat{\pi_{\mathrm{k}}} = \frac{\sum_{u} P(\mathrm{Z}_{\mathrm{u}}^{1} = \mathrm{k} \mid \mathrm{I}_{\mathrm{u}}^{1:\mathrm{T}}; \Theta^{old}) + \alpha_{\mathrm{k}} - 1}{\sum\sum P(\mathrm{Z}_{\mathrm{u}}^{1} = \mathrm{k} \mid \mathrm{I}_{\mathrm{u}}^{1:\mathrm{T}}; \Theta^{old}) + \alpha - K} \quad \text{(B8)}$$

Similarly, we can get:

$$\widehat{A_{\mathrm{jk}}} = \frac{\sum_{u}\sum_{t}^{T} P(\mathrm{Z}_{\mathrm{u}}^{t-1} = j, \mathrm{Z}_{\mathrm{u}}^{t} = k \mid \mathrm{I}_{\mathrm{u}}^{1:\mathrm{T}}; \Theta^{old}) + \alpha_{\mathrm{k}} - 1}{\sum_{u}\sum_{t}^{T}\sum_{l} P(\mathrm{Z}_{\mathrm{u}}^{t-1} = j, \mathrm{Z}_{\mathrm{u}}^{t} = l \mid \mathrm{I}_{\mathrm{u}}^{1:\mathrm{T}}; \Theta^{old}) + \alpha - K} \quad \text{(B9)}$$

$$\widehat{D}_{k,d} = \frac{\sum_{u}\sum_{t=1}^{T} p(Z_{u}^{t} = k, D_{u}^{t} = d \mid \mathrm{I}_{\mathrm{u}}^{1:\mathrm{T}}; \Theta^{old}) + \alpha_{d} - 1}{\sum_{u}\sum_{t=1}^{T}\sum_{d=1}^{M} p(Z_{u}^{t} = k, D_{u}^{t} = d \mid \mathrm{I}_{\mathrm{u}}^{1:\mathrm{T}}; \Theta^{old}) + \alpha - M} \quad \text{(B10)}$$

$$\hat{\theta}_k(I_j)=\frac{\sum_u\sum_{t=1}^{T}P(\mathrm{Z}_\mathrm{u}^t=k\mid \mathrm{I}_\mathrm{u}^{1:\mathrm{T}};\Theta^{old})\sum_j 1_i(\mathrm{I}_\mathrm{uj}^t)+\alpha_j-1}{\sum_u\sum_{t=1}^{T}P(\mathrm{Z}_\mathrm{u}^t=k\mid \mathrm{I}_\mathrm{u}^{1:\mathrm{T}};\Theta^{old})N_u^t+\alpha-\mid I\mid} \quad \text{(B11)}$$

APPENDIX C

**Maximum likelihood estimation for negative binomial distribution**

We employ Newton's method to find the approximate numerical values of parameters *r, p* in negative binomial distribution. The distribution can be expressed by:

$$f(k;r,p)=\frac{\Gamma(k+r)}{k!\Gamma(r)}p^k(1-p)^r \quad \text{(C1)}$$

The log-likelihood function is:

$$\begin{aligned}L(r,p,\lambda)&=\sum_u\sum_{t=1}^{T}\sum_k^{K}\sum_{\mathrm{d}}^{M}[p(Z_u^t=k,D_u^t=d\mid \mathrm{I}_\mathrm{u}^{1:\mathrm{T}};\Theta^{old})\log p(N_\mathrm{u}^\mathrm{t}\mid r_{k,d},p_{k,d})]\\&=\sum_u\sum_{t=1}^{T}\sum_k^{K}\sum_{\mathrm{d}}^{M}p(Z_u^t=k,D_u^t=d\mid \mathrm{I}_\mathrm{u}^{1:\mathrm{T}};\Theta^{old})[\log\Gamma(N_\mathrm{u}^\mathrm{t}+r_{k,d})-\log\Gamma(r_{k,d})\\&\quad-\log N_\mathrm{u}^\mathrm{t}!+N_\mathrm{u}^\mathrm{t}\log p_{k,d}+r_{k,d}\log(1-p_{k,d})]\end{aligned} \quad \text{(C2)}$$

We take the partial derivatives with respect to *r* and *p*, and set them equal to zero:

$$\frac{\partial L}{\partial r_{\mathrm{k},d}}=\sum_u\sum_{t=1}^{T}p(Z_u^t=k,D_u^t=d\mid \mathrm{I}_\mathrm{u}^{1:\mathrm{T}};\Theta^{old})[\Psi(r_{\mathrm{k},d}+N_\mathrm{u}^t)-\Psi(r_{\mathrm{k},d})+\log(1-\mathrm{p}_{\mathrm{k},d})] \quad \text{(C3)}$$

$$\frac{\partial L}{\partial p_{k,d}}=\sum_u\sum_{t=1}^{T}p(Z_u^t=k,D_u^t=d\mid \mathrm{I}_\mathrm{u}^{1:\mathrm{T}};\Theta^{old})[\frac{N_\mathrm{u}^t}{p_{k,d}}-\frac{r_{\mathrm{k},d}}{1-p_{k,d}}] \quad \text{(C4)}$$

where $\Psi(\mathrm{r})=\Gamma'(r)/\Gamma(\mathrm{r})$ is the digamma function. Solving the second formula (C4) for *p* gives:

$$p_{k,d}=\mathrm{E}(N_\mathrm{u}^t)/[\mathrm{E}(N_\mathrm{u}^t)+r_{\mathrm{k},d}] \quad \text{(C5)}$$

$$\text{where }\ \mathrm{E}(N_\mathrm{u}^t)=\frac{\sum_u\sum_{t=1}^{T}p(Z_u^t=k,D_u^t=d\mid \mathrm{I}_\mathrm{u}^{1:\mathrm{T}};\Theta^{old})N_\mathrm{u}^t}{\sum_u\sum_{t=1}^{T}p(Z_u^t=k,D_u^t=d\mid \mathrm{I}_\mathrm{u}^{1:\mathrm{T}};\Theta^{old})r_{\mathrm{k},d}}$$

Substituting this in the first formula (C3) gives:

$$\frac{\partial L}{\partial r_{\mathrm{k},d}}=\sum_{u}\sum_{t=1}^{T}p(Z_u^t=k,D_u^t=d\mid \mathrm{I}_{\mathrm{u}}^{1:\mathrm{T}};\Theta^{old})[\Psi(r_{\mathrm{k},d}+N_{\mathrm{u}}^t)-\Psi(r_{\mathrm{k},d})+\log(\frac{r_{\mathrm{k},d}}{\mathrm{E}(N_{\mathrm{u}}^t)+r_{\mathrm{k},d}})] \tag{C6}$$

This formula cannot be solved for $r$ in closed form. Newton's method is used to get the numerical solution.

Assume that $f(r_{k,d}) = \partial L/\partial r_{k,d}$ , we can get its derivative with respect to $r_{k,d}$.The process is repeated as:

$$r_{n+1}=r_n-f(r)/f'(r) \tag{C7}$$

until a sufficiently accurate value is reached.